\documentclass[journal=jacsat,manuscript=article]{achemso}
\usepackage[version=3]{mhchem} 
\usepackage{xcolor}
\usepackage{adjustbox}
\usepackage{subfig}

\usepackage[T1]{fontenc}
\usepackage[utf8]{inputenc}
\usepackage{verbatim}
\usepackage{graphicx}
\usepackage{rotating}
\usepackage{amsmath}
\usepackage{amssymb}
\usepackage{xr}
\usepackage{siunitx}
\usepackage{multirow}

\title{Comprehensive Study of Lithium Adsorption and Diffusion on Janus Mo/WXY (X, Y= S, Se, Te) using First Principles and Machine Learning Approaches}

\author{Gracie Chaney}
\affiliation{Department of Physics, University of Maryland Baltimore County, 1000 Hilltop Circ., Baltimore, MD 21250, USA}
\author{Akram Ibrahim}
\affiliation{Department of Physics, University of Maryland Baltimore County, 1000 Hilltop Circ., Baltimore, MD 21250, USA}
\author{Fatih Ersan}
\affiliation{Department of Physics, University of Maryland Baltimore County, 1000 Hilltop Circ., Baltimore, MD 21250, USA}
\alsoaffiliation{Department of Physics, Ayd{\i}n Adnan Menderes University, Ayd{\i}n 09100, Turkey}
\author{D. Çakır}
\affiliation{Department of Physics and Astrophysics, University of North Dakota, Grand Forks, North Dakota 58202, USA}
\author{Can Ataca}\email{ataca@umbc.edu}
\affiliation{Department of Physics, University of Maryland Baltimore County, 1000 Hilltop Circ., Baltimore, MD 21250, USA}

\begin{document}

\begin{abstract}
The structural asymmetry of two-dimensional (2D) Janus transition metal dichalcogenides (TMDs) produces internal dipole moments that result in interesting electronic properties.  These properties differ from the regular (symmetric) TMD structures that the Janus structures are derived from.  In this study, we, at first, examine adsorption and diffusion of a single Li atom on regular MX$_{2}$ and Janus MXY (M = Mo, W; XY = S, Se, Te) TMD structures at various concentrations using first principles calculations within density functional theory. Lithium adsorption energy and mobility differ on the top and bottom sides of each Janus material. The correlation between Li adsorption energy, charge transfer and bond lengths at different coverage densities are carefully examined. To gain more physical insight and prepare for future investigations of regular TMD and Janus materials, we applied a supervised machine learning (ML) model that uses cluster-wise linear regression to predict the adsorption energies of Li on top of 2D TMDs.  We developed a universal representation with few descriptors that take into account the intrinsic dipole moment and the electronic structure of regular and Janus 2D layers, the side where the adsorption takes place and the concentration dependence of adatom doping. This representation can easily be generalized to be used for other impurities and 2D layer combinations, including alloys as well.  At last, we focus on analysing these structures as possible anodes in battery applications. We conducted Li diffusion, open-circuit-voltage and storage capacity simulations. We report that Lithium atoms are found to easily migrate between transition metal (Mo, W) top sites for each considered case, and in these respects many of the examined Janus materials are comparable or superior to graphene and to regular TMDs. In addition, we report that the side with higher electronegative chalcogen atoms are suitable for Li adsorption and only MoSSe and MoSeTe can be suitable for full coverage of Li atoms on the surface. Bilayer Li adsorption was hindered due to negative open circuit voltage. Bilayer Janus structures are better suited for battery applications due to less volumetric expansion/contraction during discharging/charging process and having higher storage capacity. Janus monolayers transition from semiconducting to metallic upon adsorption of a single Li-ion, which would improve anode conductivity.  The results imply that the examined Janus structures should perform well as electrodes in Li-ion batteries.   

    \noindent
    \textcolor{red}{\textbf{KEYWORDS:} two-dimensional (2D) materials, transition metal dichalcogenide (TMD), Janus materials, Lithium-ion batteries, density functional theory (DFT), machine learning, descriptor design, adsorption energy prediction}

\end{abstract}

\begin{tocentry} 
\includegraphics{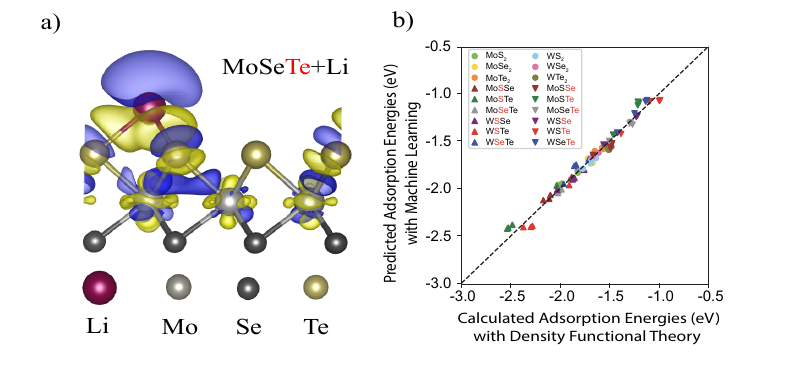}
\end{tocentry}

 \maketitle

\section{Introduction}
In recent years, much research has focused on developing renewable energy sources and storage devices.\cite{qazi}  Special consideration has gone into improving the stability/performance/capacity of the Li ion battery that may in the future be used to power hybrid/full electric vehicles, portable electronics, and so on.\cite{kim}  It is attractive as an electrochemical energy storage device because of its potentially high energy and power capacities.\cite{jansen, chen}  Two-dimensional (2D) materials are well suited as anode materials in Li-ion batteries.  With their large surface area, high electrical and thermal conductivity, and mechanical strength, 2D materials such as graphene, silicene, phosphorene and various MXs\cite{zhang_1, zhang_2} are superior anode materials to their bulk counterparts.\cite{mukherjee} Intensive research has focused on 2D transition metal dichalcogenides (TMDs).\cite{ataca}  

TMD monolayers are semiconductors consisting of transition metal atoms (ie. Mo, W) between layers of chalcogenide atoms (ie. S, Se, Te).  Both bulk and 2D TMDs are very important to electronic and optical applications.\cite{karmodak, aguilera-granja} Their monolayer forms are intriguing because of the properties that arise due to surface effects.\cite{choi}  For instance, bulk TMD materials have indirect band gaps, while their monolayer versions have direct band gaps in the visible or infrared ranges.  MoS$_2$ and WS$_2$ monolayers, for example, have direct band gaps of 1.8 eV and 1.5 eV respectively.\cite{choi} In contrast, graphene has a very high carrier mobility due to its linear dispersion relation at its Kappa point\cite{katsenelson}; however, it is not a suitable material for devices such as field effect transistors, because it lacks a band gap.\cite{luo} TMDs have tunable band gaps that allow them to transition between metallic and semiconducting forms.  Thus, TMDs are versatile materials that can be used for transistors, solar cells, integrated circuits and so on.\cite{luo, zhang_3} TMDs are also strong candidates for anode materials in lithium-ion batteries.  Such anode materials exhibit high lithium-ion mobility as they have large interlayer distances and interact only weakly with the ions.\cite{xu} This means that the TMD anodes are not subject to the rapid volume changes and capacity decreases that Si anodes experience upon intercalation of Li-ions.\cite{zhao, ko}  Ersan \textit{et al.}\cite{ersan} suggests that TMDs are superior anode materials to graphene and silicene, which are the two-dimensional versions of graphite and silicon.  According to their work, the energy barrier for a single Li atom migrating parallel across a TMD monolayer is between 0.15 eV and 0.28 eV,
whereas the lowest energy barriers for a Li atom on unstrained and pristine graphene and silicene sheets are 0.23 eV (for hollow-bridge-hollow site path)\cite{zhong} and 0.27 eV (for hollow-top-hollow site path)\cite{wang}.  This means that the Li needs less energy to diffuse across most of the TMD monolayers than on the more conventional anode materials.

Researchers are also exploring a variation of the TMD, called the Janus monolayer.\cite{ersan}  Janus monolayers are derived from regular TMDs that have had one of its chalcogenide layers removed and replaced with atoms of another chalcogen atom. Recently, Janus TMD monolayers (MoSSe) have been experimentally synthesized. Zang \textit{et al.} \cite{zhang}, synthesized 2D MoSSe from MoSe$_2$ by removing the top layer of Se from the material and substituting S atoms.  The same year, Lu \textit{et al.} \cite{lu} synthesized MoSSe by starting from MoS$_2$ and replacing one it its S layers with Se atoms. Structural symmetry is key to determining a material's electronic properties.  Janus monolayers break out-of-plane mirror symmetry thus inducing an electric dipole moment between the chalcogenide layers.\cite{shi, tao} This intrinsic dipole moment influences the Janus TMDs' electronic, magnetic, and adsorption properties.  For instance, it can induce piezoelectricity in the materials.\cite{tao} 

After Janus TMD monolayers were successfully synthesized, theorists began to simulate these and other Janus structures to understand and predict their properties.  For instance, Shang \textit{et al.}\cite{shang} used Density Functional Theory (DFT) to study MoSSe’s potential as an anode material in Li-ion batteries.  They looked at how the Li-ions diffuse across each side of the Janus monolayer.  Around the same time Xiong \textit{et al.}\cite{xiong} used molecular dynamics simulations to predict that a free standing MoSSe monolayer will spontaneously curl into a nanotube if it's circumference is 33 nm or larger.  This occurs because of the asymmetry of the Janus monolayer that results in different bond lengths and angles for the S and Se sides.  Using DFT, Shi \textit{et al.}\cite{shi} investigated the mechanical and electronic properties of Janus monolayers MXY (M = Ti, Zr, Hf, V, Nb, Ta, Cr, Mo, W; X/Y = S, Se, Te) with 2H and 1T phases.  Specifically, they calculated the elastic constants and electronic structure of these materials. Cheng \textit{et al.}\cite{cheng} conducted stability analysis including phonon dispersions of MXY (M=Mo,W and X/Y=S,Se and Te) and reported that they are all stable. It is known that applying strain to a TMD can effect its electronic properties, and can even induce a transition from a semiconducting to a metallic state.  Liu \textit{et al.}\cite{liu} showed that applying positive or negative strain to Janus TMDs MXY (M = Mo, W; X/Y = S, Se, Te) changes the materials' band gaps. For example, applying tensile strain to MoSSe decreases it's band gap as well as induces a transition from direct to indirect band gap.  Thus, applying strain to MoSSe can increase its light absorption range and reduce the recombination rate of photo-generated carriers.  Thus, it appears that MoSSe and perhaps other Janus TMDs are potential materials for water-splitting photocatalysts.\cite{ma,ju} 

In this study, we seek to understand the Janus monolayers' potential as anode materials in Li-ion batteries. At first, we study a single Li adsorption at varying concentrations and then examine the mobility of the Li ions as they diffuse parallel to the monolayers.  This is not a trivial problem as it is not well understood how the dipole moment affects diffusion.  We use DFT to predict the migration paths and diffusion coefficients of a single Li ion as it diffuses across various Janus monolayers.  We do so by calculating the adsorption energies of the Li ion at multiple sites of the monolayer.  We then use this data for the nudged elastic band (NEB) method.  Specifically, we study Li diffusion on all of the combinations of two transition metals (Mo and W), and three chalcogens.(S, Se, Te).  For all of the structures, we find that the Li ion prefers to diffuse between metal top sites, passing over a hollow site in the process.  Also, we acquire activation energy plots for both sides of each Janus structure and calculate the diffusion coefficients, revealing that the Li ions will more easily diffuse on these monolayers than on traditional TMDs and other anode materials. Secondly, we focus on stability and performance of these materials. We calculate formation energies of Li adsorption, open-circuit voltage at varying Li coverage and storage capacities including multilayer Li adsorption and multilayer Janus structures. In addition, we report how thicknesses of the Janus structures are changing upon Li adsorption and provide ways to avoid volume expansion upon charging/discharging. Thus, we believe that Janus TMDs are superior candidates for anodes of Li ion batteries.

We go a step further and employ a supervised machine learning model for the adsorption energies acquired with DFT.  We follow a similar formalism to that introduced by Dou \emph{et al.}\cite{dou} and rely on an initial theoretical analysis of the energetics of the adsorption process that simplifies the adsorption energy to few energy terms which then serve as the descriptors for our linear model with a low-dimensional feature space.
Dou \emph{et al.} used the energy of the lowest unoccupied state $E
_{LUS}$, the energy of the conduction band minimum (CBM) with respect to vacuum, as the main descriptor for their linear regression model.  They used this model to predict the adsorption energies of different alkali atoms on regular TMD monolayers.  In this work, we adopt a similar approach while extending the scope of our predictive model to include adsorption on both regular TMDs and the different sides of Janus TMDs.  In addition, we take various levels of Li coverage into account, from full coverage to dilute.  In order to achieve that, we use a cluster-wise linear regression (CLR) model.\cite{1250952}  To disentangle the heterogeneity in a population, CLR divides the dataset into K (>1) subsets (clusters) such that each cluster represents a homogeneous subpopulation. Then K linear regression models are applied to each subset separately resulting in a much smaller residue error than without clustering.  We have clustered our dataset simply based on  supercell size, which has proven to be a valid assumption.  Then we applied an ordinary least squares (OLS) linear regression within each cluster.  In this work, we will show that our model, which is based on only three main descriptors, is efficient and provides very good prediction results of adsorption energies for Li on both regular and Janus TMD structures at different coverage ratios.

\section{Computational Methods}

Our calculations are based on the first-principles plane wave method within density functional theory (DFT) using pseudopotentials supplied by Vienna Ab initio Simulation Package (VASP).\cite{vasp}  We approximate the exchange correlation functional with the Perdew Burke Ernzerhof (PBE) approach of the generalized gradient approximation (GGA).\cite{gga}  We perform self-consistent field (scf) potential and ionic relaxation calculations with VASP for $1 \times 1 \times 1$, $2 \times 2 \times 1$, $3 \times 3 \times 1$, and $4 \times 4 \times 1$ supercells.  In these calculations, the Brillouin Zone (BZ) is sampled using $(24 \times 24 \times 1)$, $(12 \times 12 \times 1)$, $(8 \times 8 \times 1)$ and $(6 \times 6 \times 1)$ Monkhorst-Pack special k-points meshes that correspond to the primitive cell and  three supercell sizes. We use a plane wave basis set with kinetic energy cutoff of $\hbar^2{(\textbf{k+G})}^2$/2m = 500 eV for the regular and Janus TMD structures.  We set the energy convergence criterion between two successive iterations to be 10$^{-5}$ eV, and kept the pressure less than 1.00 kBar.  The Fermi surface is Gaussian smeared by 0.01 eV. To avoid interlayer interaction, two MXY layers are separated by a 20 \si{\angstrom} vacuum region. Spin-orbit interactions are considered for local potential simulations for side-dependent work function calculation of Janus structures. The rest of the adsorption and diffusion simulations are spin-polarized, and van der Waals (vdW) corrections were used.\cite{grimme} Also, we performed dipole moment calculations on the bare primitive cells of all regular and Janus TMDs. After geometry relaxation, we performed Bader analysis and band structure generation for all materials of every supercell size. 

We calculate the adsorption energies of the Li atom on multiple lattice sites for both sides of each monolayer. This gives us the start and end locations of the migrating Li atom, which are the metal top sites for all the configurations.  We used the climbing image nudged elastic band method (CI-NEB) to find the minimum energy path between the initial and final positions of a Li atom hopping between adjacent metal top sites.\cite{henkelman}  NEB calculations are conducted in $3 \times 3 \times 1$ supercells and 11 images are taken between neighboring ground state sites. From the NEB calculations, we acquired the diffusion energy barriers, E$_a$, that must be overcome for a Li atom to hop between nearest neighbor lattice sites.  

\section{Results and Discussion} 

\subsection{Bare Structures}

We begin by analysing the bare regular and Janus TMD primitive cells.  In Tab. \ref{tab:table1}, we provide the optimized lattice constants, charges on each atom species, dipole moment, and cohesive energy of each examined structure.  Our results show that the lattice constant of a Janus material is determined by those of the regular TMD it was derived from, and depends on which species were removed or added.  In fact, all the lattice constants for Janus MXY structures are the average of the constituent MX$_2$ monolayers.  The conversion from regular TMD to Janus can cause either an expansion or shrinking of the lattice constant, the extent of which depends on the atomic radii of the chalcogen atoms.  For instance when a structure transforms from MoS$_2$ to MoSSe, the lattice constant increases by only 0.06 \AA{} (1.88 \%). Se has a larger atomic radius than S, so the structure expands after one of its S layers is replaced with Se.  Removing another S layer and replacing it with Se (MoSe$_2$), causes the lattice constant to expand 0.07 \AA{} (2.15 \%)  more.  Both of these expansions are slight, because the difference in atomic radii between S and Se is not large enough to cause a drastic change of the lattice constant. In contrast, the radius of Te is more than twice that of S.  After MoS$_2$ transforms to MoSTe, the lattice constant expands by 5.10 \%.  After transforming from MoTe$_2$ to MoSeTe, the lattice constant decreases by 5.52 \%.  The same pattern exists for W based TMDs as well. 

\begin{table}[h!]
    \centering
    \caption{Regular TMD and Janus primitive cells; MXY hexagonal lattice constant ($|a|=|b|$) in \AA{}, charge of metal ($\Delta \rho_{M}$) and chalcogenide atoms in electrons, $e$ ($\Delta \rho_{X}$, $\Delta \rho_{Y}$), dipole moment ($\mu$) in e\AA{}, cohesive energy per atom in eV ($E_{coh}$).}
    \begin{tabular}{c|c|c|c|c|c|c|c}
        \textbf{TMD} & \textbf{a(\AA{})} & \boldmath$\Delta \rho_{M} (e)$ & \boldmath$\Delta \rho_{X} (e)$ & \boldmath$\Delta \rho_{Y} (e)$ & \boldmath$\mu (D)$ & \boldmath$E_{coh} (eV)$ \\
         \hline
         MoS$_2$ & 3.2 & 1.07 & -0.54 & - & - & -5.34 \\
         MoSe$_2$ & 3.33 & 0.82 & -0.41 & - & - & -4.86 \\
         MoTe$_2$ & 3.56 & 0.48 & -0.24 & - & - & -4.57 \\
         WS$_2$ & 3.19 & 1.15 & -0.58 & - & - & -6.09 \\
         WSe$_2$ & 3.34 & 0.91 & -0.46 & - & - & -5.56 \\
         WTe$_2$ & 3.58 & 0.52 & -0.26 & - & - & -5.02 \\
         MoSeTe & 3.43 & 0.56 & -0.34 & -0.22 & 0.17 & -4.59 \\
         MoSSe & 3.26 & 0.91 & -0.55 & -0.36 & 0.10 & -5.10 \\
         MoSTe & 3.36 & 0.78 & -0.57 & -0.21 & 0.22 & -4.79 \\
         WSeTe & 3.43 & 0.71 & -0.47 & -0.25 & 0.64 & -5.26 \\
         WSSe & 3.25 & 1.03 & -0.62 & -0.42 & 0.14 & -5.82 \\
         WSTe & 3.36 & 0.87 & -0.64 & -0.23 & 0.88 & -5.47 \\
    \end{tabular}
    \label{tab:table1}
\end{table}

Table \ref{tab:table1} includes the charge transfer of each atomic species.  The data implies that the transition metal atom (M=Mo,W) donates charge to the surrounding chalcogenide atoms (X/Y=S,Se,Te) for both regular and Janus TMDs.  The charges are not distributed evenly in the Janus materials, however. In MoSeTe, for instance, more charge is transferred to the Se layer than to the Te layer because the former is more electronegative than the latter.  The order of electronegativity is as follows: Te < Se < S.  Fundamentally, the reason for this asymmetric charge transfer in Janus TMDs is that these materials break mirror symmetry by having two chalcogenide layers of different species.  This asymmetry is what gives Janus materials their unique properties. In particular, an intrinsic polarization develops that points vertically from one chalcogenide layer to the other.  These dipole moments, recorded in Tab. \ref{tab:table1}, determine most of the properties of the Janus structures, such as adsorption of adatoms on their surfaces.  Notice that for both the Mo and W structures, those Janus TMDs with S and Te have the highest dipole moments.  This is due to the high electronegativity contrast between S and Te.  Similarly, S and Se have almost the same electronegativity value, so MoSSe and WSSe have lower dipole moments compared to the other structures.

Finally, we calculated the cohesive energy of the regular and Janus TMDs to compare their mechanical stabilities. We define cohesive energy as

\begin{equation}
    E_{coh} = \frac{E_{MX_{2}/MXY} - E_{M} - E_{X} - E_{Y}}{3}
\end{equation}

where $E_{MX_{2}/MXY}$, $E_{M}$, and $E_{X/Y}$ are the total energies of the primitive cell of regular or Janus TMDs, a single transition metal atom, and a chalcogen atom in vacuum.  $E_{X}$ = $E_{Y}$ for regular TMDs. The results are given in Tab. \ref{tab:table1}. As expected, we found that cohesive energy decreases as lattice constant increases.  We also found that cohesive energy depends on the electronegativity of the chalcogenide atoms. Physically, this is because more charge is transferred from the transition metal to the more electronegative chalcogens, acting as "glue" for the structures.  Of the regular TMDs, MoS$_2$ and WS$_2$ have the largest cohesive energies of the Mo and W groups, respectively.  Of the Janus TMDs, MoSSe and WSSe have the largest values of the Mo and W groups, respectively.  This result is due to the combination of S and Se, the more electronegative elements.

\subsection{Lithiated Structures}

\begin{figure}[!ht]
\centering
\includegraphics[width=8cm]{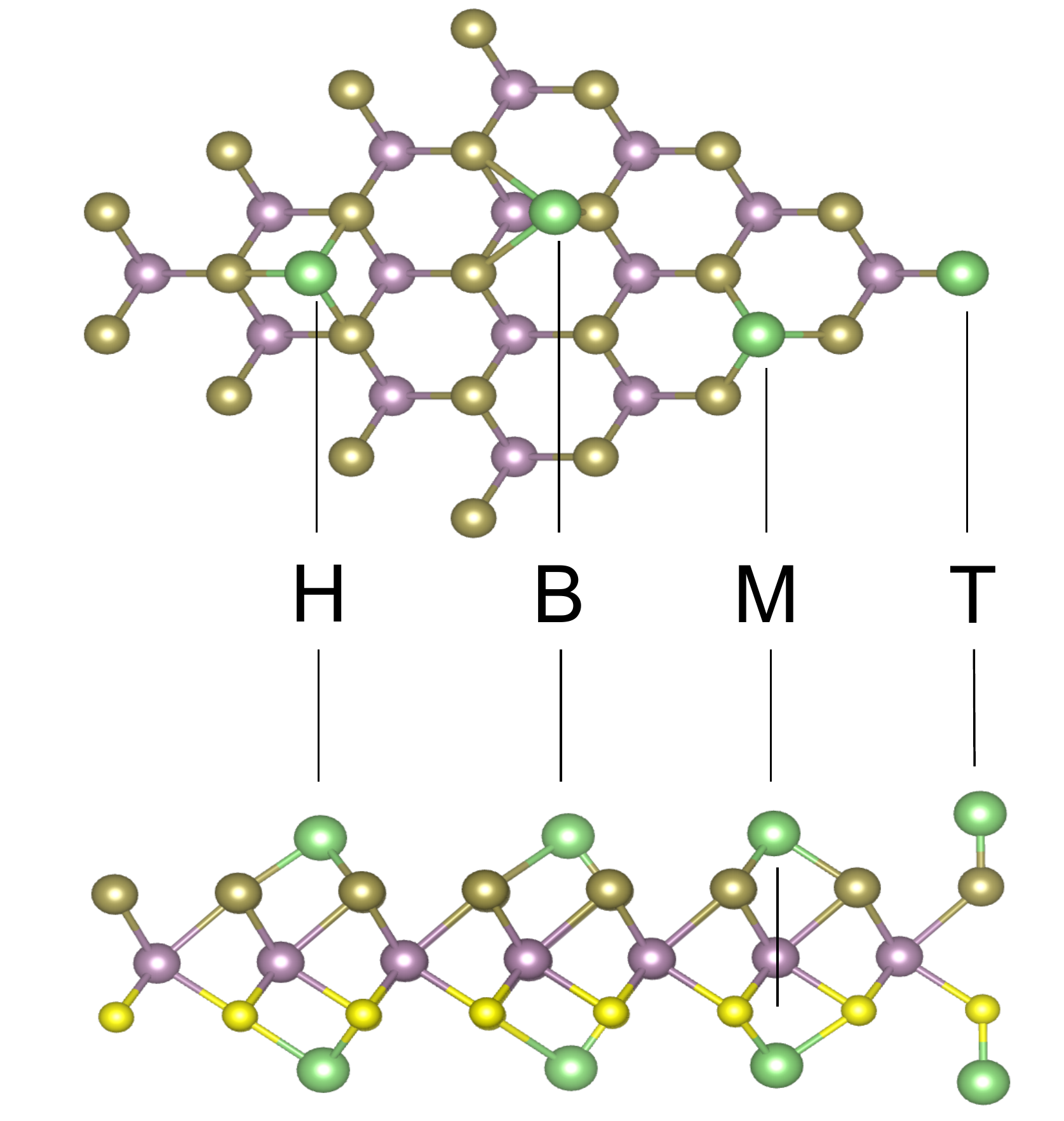}
\caption{\label{fig1-adsorption} Top and side views of Janus monolayer with Li (green balls) bound to various adsorption sites.  From left to right, the adsorption sites are as follows: Hollow (H), Bridge (B), top-metal (M), and top-chalcogen (T).  The monolayer MXY, consists of transition metal (purple balls) M: Mo, W and chalcogenides (brown, yellow balls) X,Y: S, Se, Te.}
\end{figure}

Now we examine the effects of adding Li to regular and Janus TMDs. To study how adsorption of a single Li atom on Janus structures modifies properties of and compares to that on regular TMDs, we construct supercells.  We construct $2 \times 2 \times 1$, $3 \times 3 \times 1$, and $4 \times 4 \times 1$ supercells for each Janus structure examined in order to test how coverage affects our acquired energies.  We consider four adsorption sites for single Li atom adsorption, denoted in Figure \ref{fig1-adsorption}. These sites are as follows: (i) Hollow site (H) at the center of the hexagon; (ii) Top site of the chalcogen atom (T); (iii) Top site of the transition metal atom (M); and (iv) Bridge site (B) at the middle of the bond between nearest M and (X,Y) atoms.  Note that Janus monolayers are asymmetrical, so we examine these adsorption sites for both the top and bottom layers of each material.  Initially, the Li ion is located at 2 Å above the atoms in the vacuum direction on related sites.  All of the calculations were done on optimized structures that were found upon geometric relaxation.  The adsorption energy is given as

\begin{equation}
    E_{d} = E_{tot} - E_{Li} - E_{bare}
\end{equation}
where $E_{tot}$, $E_{Li}$, and $E_{bare}$ are the total energies of the monolayer and adsorbed Li atom system, the isolated Li atom spin-polarized, and the bare monolayer in corresponding supercell sizes. According to this definition, a negative value for $E_d$ corresponds to binding of the adatom.

\begin{table}[h!]
\centering 
    \caption{Top rows of each monolayer subsection features adsorption energies of Li on regular /Janus TMD monolayers (in eV).  Second rows feature formation energies (in eV).  Third rows feature positive charge on Li after adsorption (1 - e$^{-}$ lost by Li (in $e$).  Bottoms rows feature average Li-X,Y distance (in \AA{}).  Data for all TMD/Janus monolayers for coverage ratios for primitive cell [$\theta$(1/3)], $2 \times 2 \times 1$ [$\theta$(1/12)], $3 \times 3 \times 1$ [$\theta$(1/27)], and $4 \times 4 \times 1$ [$\theta$(1/48)] supercells are tabulated. The side where lithiation occurs and adsorption site is also reported.}
    
\footnotesize
\tiny
    \begin{tabular}{c|c|c|c|c|c|c}

        \textbf{TMD} & \textbf{Side} & \textbf{Ads Site} & \boldmath$\theta$\textbf{(1/3)} & \boldmath$\theta$\textbf{(1/12)} & \boldmath$\theta$\textbf{(1/27)}& \boldmath$\theta$\textbf{(1/48)}\\
        \hline
       
         MoS$_2$ & - & M & -1.824 & -1.894 & -1.986 & -2.024 \\ 
         &    &           & -0.110 & -0.180 & -0.272 & -0.310  \\ 
         &    &           & 0.76 & 0.857 & 0.870 & 0.874 \\ 
         &    &           & 2.475 & 2.342 & 2.374 & 2.385 \\ \hline
         MoSe$_2$ & - & M & -1.724 & -1.674 & -1.657 & -1.674 \\
         &    &           & -0.010 & 0.040 & 0.057 & 0.040  \\
         &    &           & 0.759 & 0.854 & 0.864 & 0.866 \\ 
         &    &           & 2.653 & 2.463 & 2.484 & 2.494 \\ \hline
         MoTe$_2$ & - & M & -1.580 & -1.654 & -1.544 & -1.564 \\
         &    &           & 0.134 & 0.060 & 0.170 & 0.150  \\
         &    &           & 0.752 & 0.851 & 0.859 & 0.860 \\ 
         &    &           & 2.853 & 2.686 & 2.698 & 2.708 \\ \hline
         WS$_2$ & - & M & -1.695 & -1.774 & -1.641 & -1.688 \\
         &    &           & 0.019 & -0.060 & 0.073 & 0.026  \\
         &    &           & 0.754 & 0.856 & 0.865 & 0.868 \\ 
         &    &           & 2.500 & 2.328 & 2.359 & 2.363 \\ \hline
         WSe$_2$ & - & M & -1.614 & -1.676 & -1.486 & -1.516 \\
         &    &           & 0.100 & 0.34 & 0.228 & 0.198  \\
         &    &           & 0.755 & 0.851 & 0.858 & 0.860 \\ 
         &    &           & 2.669 & 2.439 & 2.464 & 2.468 \\ \hline
         WTe$_2$ & - & M & -1.492 & -1.511 & -1.485 & -1.522 \\
         &    &           & 0.222 & 0.204 & 0.229 & 0.192  \\
         &    &           & 0.748 & 0.849 & 0.856 & 0.859 \\ 
         &    &           & 2.880 & 2.666 & 2.675 & 2.682 \\ \hline
         
         MoSeTe & Se & M & -1.828 & -2.027 & -1.990 & -2.021 \\
         &    &           & -0.114 & -0.313 & -0.276 & -0.307  \\
         &    &           & 0.764 & 0.852 & 0.861 & 0.862 \\ 
         &    &           & 2.611 & 2.467 & 2.491 & 2.497 \\ \hline
         MoSeTe & Te & M & -1.508 & -1.272 & -1.291 & -1.301 \\
         &    &           & 0.205 & 0.441 & 0.422 & 0.413  \\
         &    &           & 0.741 & 0.852 & 0.862 & 0.865 \\ 
         &    &           & 2.903 & 2.657 & 2.726 & 2.736 \\ \hline
         MoSSe & S & M & -1.887 & -2.102 & -2.114 & -2.173 \\
         &    &           & -0.173 & -0.388 & -0.400 & -0.459  \\
         &    &           & 0.764 & 0.856 & 0.867 & 0.870 \\ 
         &    &           & 2.462 & 2.346 & 2.374 & 2.508 \\ \hline
         MoSSe & Se & M & -1.672 & -1.493 & -1.478 & -1.504 \\
         &    &           & 0.041 & 0.221 & 0.235 & 0.209   \\
         &    &           & 0.752 & 0.855 & 0.864 & 0.870 \\ 
         &    &           & 2.667 & 2.470 & 2.499 & 2.508 \\ \hline
         MoSTe & S & M & -2.038 & -2.486 & -2.527 & -2.537 \\
         &    &           & -0.325 & -0.773 & -0.813 & -0.823  \\
         &    &           & 0.770 & 0.857 & 0.866 & 0.868 \\ 
         &    &           & 2.469 & 2.381 & 2.397 & 2.405 \\ \hline
         MoSTe & Te & M & -1.469 & -1.219 & -1.219 & -1.206 \\
         &    &           & 0.245 & 0.494 & 0.495 & 0.508  \\
         &    &           & 0.738 & 0.850 & 0.863 & 0.866 \\ 
         &    &           & 2.917 & 2.726 & 2.748 & 2.761 \\ \hline
         
         WSeTe & Se & M & -1.714 & -1.968 & -1.842 & -1.861 \\
         &    &           & -0.001 & -0.255 & -0.128 & -0.147  \\
         &    &           & 0.761 & 0.850 & 0.857 & 0.857 \\ 
         &    &           & 2.624 & 2.467 & 2.477 & 2.481 \\ \hline
         WSeTe & Te & M & -1.422 & -1.252 & -1.126 & -1.140 \\
         &    &           & 0.292 & 0.462 & 0.587 & 0.573  \\
         &    &           & 0.741 & 0.852 & 0.857 & 0.859 \\ 
         &    &           & 2.924 & 2.809 & 2.707 & 2.713 \\ \hline
         WSSe & S & M & -1.760 & -2.033 & -1.858 & -1.877 \\
         &    &           & -0.046 & -0.319 & -0.145 & -0.163  \\
         &    &           & 0.755 & 0.856 & 0.863 & 0.867 \\ 
         &    &           & 2.486 & 2.346 & 2.364 & 2.377 \\ \hline
         WSSe & Se & M & -1.565 & -1.426 & -1.222 & -1.235 \\
         &    &           & 0.149 & 0.287 & 0.492 & 0.479  \\
         &    &           & 0.750 & 0.852 & 0.855 & 0.859 \\ 
         &    &           & 2.697 & 2.430 & 2.477 & 2.474 \\ \hline
         WSTe & S & M & -1.911 & -2.375 & -2.283 & -2.299 \\
         &    &           & -0.198 & -0.662 & -0.569 & -0.585  \\
         &    &           & 0.761 & 0.858 & 0.866 & 0.867 \\ 
         &    &           & 2.457 & 2.393 & 2.399 & 2.403 \\ \hline
         WSTe & Te & M & -1.388 & -1.099 & -0.994 & -1.001 \\
         &    &           & 0.326 & 0.615 & 0.719 & 0.712  \\
         &    &           & 0.750 & 0.855 & 0.855 & 0.857 \\ 
         &    &           & 2.951 & 2.691 & 2.738 & 2.750 \\ 
    \end{tabular}
    \label{tab:table2}
\end{table}

\begin{figure}
\includegraphics[width=8.5cm]{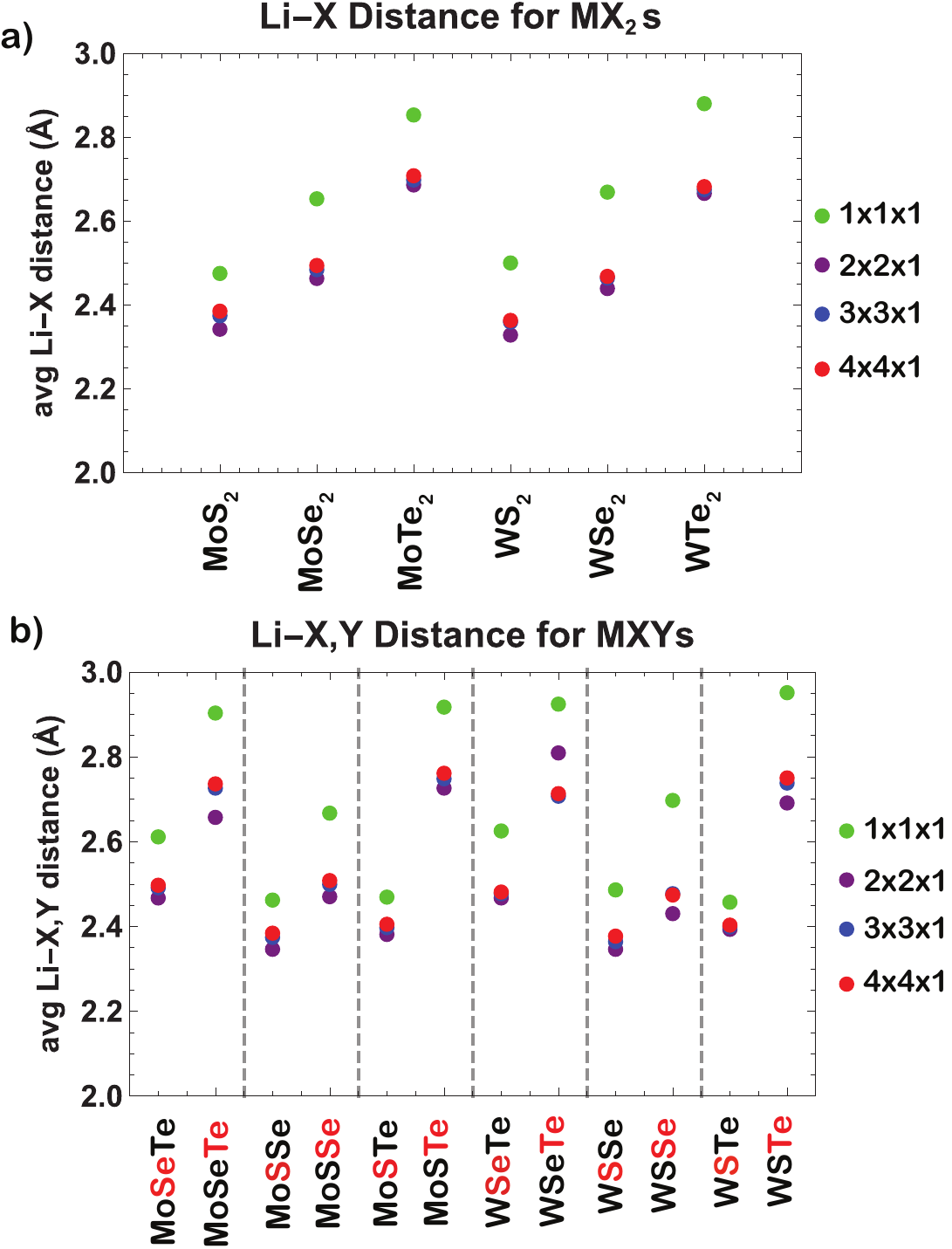}
\caption{Li-X,Y distances for Li over metal-top sites of all supercell sizes. Features a) regular TMDs and b) top and bottom sides of Janus structures.  Notice that the Li is farthest away from the $1 \times 1 \times 1$ supercells, to which Li donates the least charge. Red coloring on the chalcogen atom of Janus structures on x-axis labeling indicates which side Li atom is adsorbed at.}
\label{fig:distance}
\end{figure}

\begin{figure}
\includegraphics[width=8.5cm]{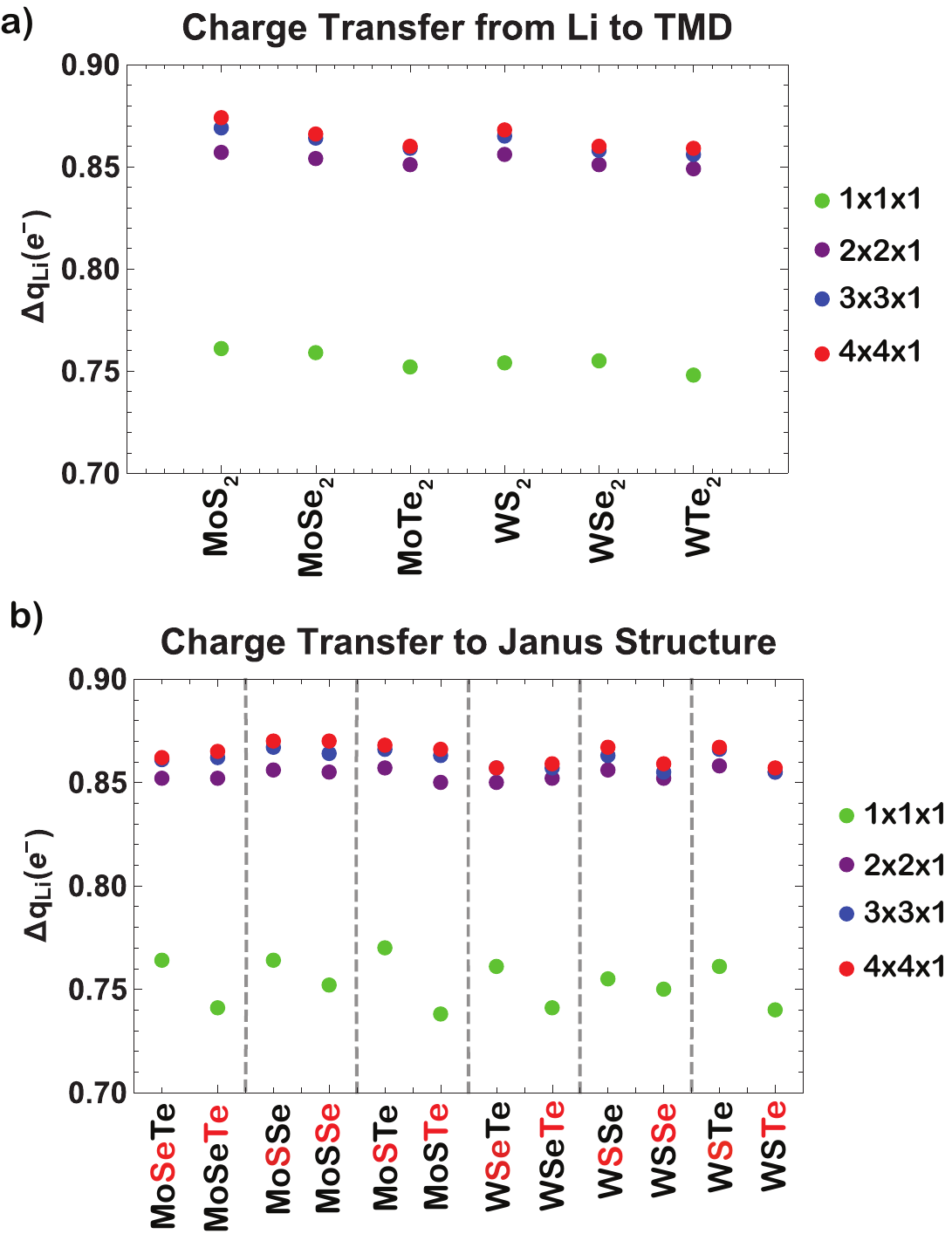}
\caption{Charge transferred from Li atom to regular/Janus TMD monolayers, over metal-top sites of all supercell sizes.  Features a) regular TMDs and b) top and bottom sides of Janus structures. Red coloring on the chalcogen atom of Janus structures on x-axis labeling indicates which side Li atom is adsorbed at.}
\label{fig:charge}
\end{figure}

Table \ref{tab:table2} reports the adsorption energies for the $1 \times 1 \times 1$, $2 \times 2 \times 1$, $3 \times 3 \times 1$, and $4 \times 4 \times 1$ cases, which are labeled in terms of coverage ratios as $\theta$(1/3), $\theta$(1/12), $\theta$(1/27), and $\theta$(1/48) respectively.  Only the highest adsorption energies are given along with the lattice site where such adsorption occurs (the top-metal site). Since charge transfer between adatom and lattice surface is part of the adsorption process, the charge of the Li atom after adsorption is also tabulated in Tab. \ref{tab:table2}. Finally, the average distance of the Li from the nearest chalcogenide atoms is included in Tab. \ref{tab:table2}, since stronger adsorption tends to lead to shorter bond lengths between the adatom and the lattice.  In order to interpret the data easier, Figures \ref{fig:distance}, \ref{fig:charge} and \ref{fig:ads} display the average Li-X,Y distance, charge transferred from the Li to the monolayers, and the adsorption energies in terms of coverage of both the regular and Janus TMDs.

Figure \ref{fig:distance} features the average Li-chalcogenide distances above the most energetically favorable adsorption sites (on top of the transition metals), found by averaging the Li-X,Y bond-lengths for the three closest X,Y atoms at different coverage concentrations. As a general trend, the average bond length increases with increasing atomic number of chalcogen atom. This is most likely due to the fact that electronegativity of chalcogen atoms increases as the atomic number of chalcogen atoms decreases, which results in an increase in the charge donation from Li to chalcogen layer. In addition, as the atomic number of chalcogen atom decreases, the atomic radius also decreases. This results in a higher binding energy as the bond length decreases. When we compare the average Li-chalcogen atom distances of regular (Fig. \ref{fig:distance} (a)) and Janus TMDs (Fig. \ref{fig:distance}(b)), we see that depending on what chalcogen atoms are present at Janus TMD surfaces, the average bond lengths are close to that of the regular TMD cases. For the dense coverage cases where Li atoms are adsorbed to every primitive cell of a regular or Janus TMD, the average bond lengths are the largest. This is due to charge saturation which we will discuss in the next paragraphs. For all of the adsorption cases, the average bond distances on adsorption of Li on $3 \times 3 \times 1$ and $4 \times 4 \times 1$ supercells are very close (less than $1\%$ difference). This enables us to state that the bond-lengths converge to the dilute case around a $3 \times 3 \times 1$ supercell size. We will have a detailed analysis why the bond lengths in $2 \times 2 \times 1$ supercells  are shorter in Section D.

In every examined case the adsorbed Li atom transfers charge to the monolayer.  Note that most of the charge is transferred to the chalcogenide atoms closest to the Li atom. Fig. \ref{fig:charge} (a) and (b) feature the charge transferred to the lattice, for various supercell sizes of regular and Janus TMD materials, respectively.  In regards to the regular TMDs: as the electronegativity of the chalcogen atom increases, the Li adatom loses more negative charge and thus takes on a more positive character.  Also notice that as supercell size increases and Li concentration becomes more dilute, the charge values converge to the same values for a Li atom isolated from other Li atoms on the lattice. 

In fact, we believe that the reason why the charge transfer values for the primitive cell are smaller than for the supercells is because of the strong Coulomb repulsion between neighboring Li adatoms due to relatively small distances ($\sim 3.5$ \AA{}) between them. This also results in significant reduction in the Li adsorption energies which we will discuss in the Section D. The effect of Coulomb repulsion diminishes around $3 \times 3 \times 1$ supercell dimensions.

\begin{figure}
\includegraphics[width=8cm]{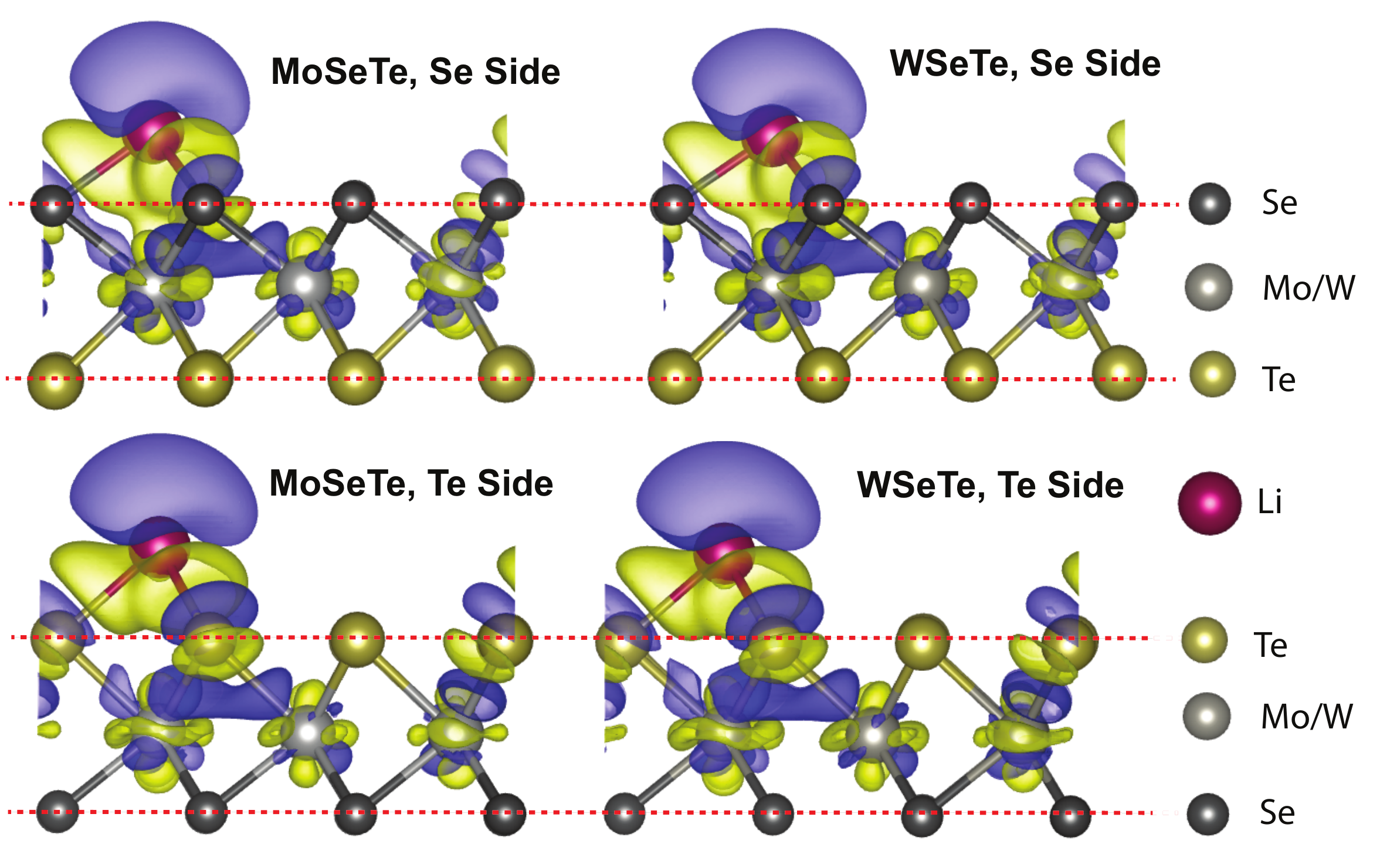}
\caption{Charge difference plots of Li adsorbed on the top and bottom sides of (Mo, W)SeTe. Yellow for charge accumulation and blue for charge depletion regions. Isosurface values for charge accumulation and depletion on Se side are taken as 0.01 and 0.008 e/\AA{}$^3$, respectively. Isosurface values for charge accumulation and depletion on Te side are taken as 0.008 and 0.004 e/\AA{}$^3$, respectively. The same trends occur for all examined Janus monolayers. Red dashed lines indicate the chalcogen layers. Coloring of atoms are indicated in the figure. }
\label{fig:chgdiff}
\end{figure}

Similar observations can be made for Janus TMD monolayers as well. Depending on the side of Janus TMD where the Li atom is adsorbed, the charge transfer values are similar to the regular TMD cases and follow similar trends with coverage as well. The intrinsic strain on the Janus TMD surfaces do not affect the charge transferred from the Li adatom. Take the case of a Li atom adsorbed to the Se side of a $3 \times 3 \times 1$ (Mo, W)SeTe structures, as indicated in the charge density plots of Fig. \ref{fig:chgdiff}.  Most of the charge is given to the three Se atoms closest to the Li, which each form Li-Se bonds of $\sim 2.49$ \AA{}. The average Bader charge delivered to them is $\sim 0.24 e^-$ each. (There is a minor charge transfer to second nearest neighbor atoms as well. See the charge depletion/ accumulation on the transition metal atoms on Fig. \ref{fig:chgdiff}) In the case of a Li atom adsorbed to the Te side of such a $3 \times 3 \times 1$ (Mo, W)SeTe structure, the Li atom again transfers most of its charge to the closest Te atoms.  The average Bader charge delivered to each of these three Te atoms is $\sim 0.23 e^{-}$, and they each form Li-Te bonds of $\sim 2.73$ \AA{}.

\begin{figure}[!ht]
\includegraphics[width=8.5cm]{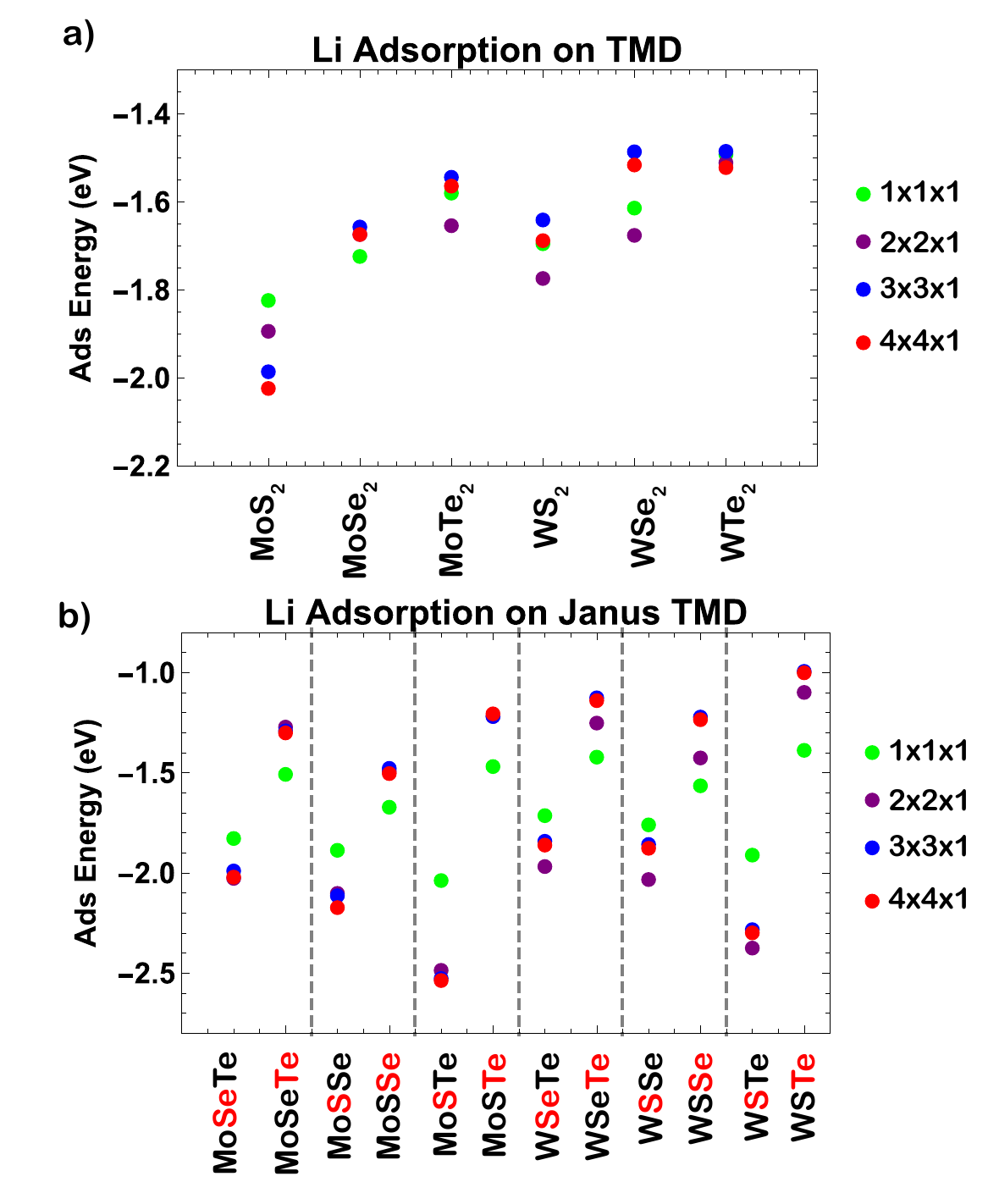}
\caption{Adsorption energies for Li atom over metal-top sites of all supercell sizes.  Features a) Mo and W based TMDs and b) top and bottom sides of Janus structures. Red coloring on the chalcogen atom of Janus structures on x-axis labeling indicates which side Li atom is adsorbed at.}
\label{fig:ads}
\end{figure}

Figure \ref{fig:ads} displays the adsorption energies in terms of coverage of both the regular (a) and Janus TMDs (b), so that the data is easier to interpret. Based on our previous discussions about Li-chalcogen atom bond length and charge transferred to Li adatom, we conclude that for regular monolayers, as the Li-X bond length decreases and Li transfers more charge to the monolayer, the binding energy of the Li adatom increases. There are three factors influencing the binding of Li adatoms. These include: actual Li-monolayer ionic bonding, Li-Li cohesion and Li-Li Coulombic repulsion. Last two interactions are playing a dominant role in understanding the binding at higher concentrations, but they decay rapidly as Li-Li distances increase which result in minor effects on dilute doping. (See Section D. Similar to the previous observations, $3 \times 3 \times 1$ supercell can be taken as a dilute doping concentration due to energy difference.) 

Overall, the Li atom acquires a higher adsorption energy when bound to the sides with the lower atomic number chalcogen (top side) rather than to the sides with the higher atomic number chalcogen (bottom side).  Most likely, this is because the top layer chalcogen atoms have electronegativities farther from that of Li than the bottom layer atoms.  On the Pauling scale, the electronegativity differences ($\Delta$EN) are as follows: 1.6 for S-Li, 1.57 for Se-Li, and 1.12 for Te-Li.  As $\Delta$EN increases, more charge flows from the Li atom to the monolayer. When binding energies of individual sides are compared with their regular TMD counterparts, they depend strongly on the change of the lattice parameter. If the lattice parameter of underlying monolayer increases (ex. comparing Li adsorption on S side of MoSSe and MoS$_2$), Li binding energy increases due to reduction in Li-Li cohesion and Coulombic repulsion. This trend reverses if the lattice parameter of underlying monolayer decreases.

Also notice that in Fig. \ref{fig:ads}(b) the adsorption energies for the $1 \times 1 \times 1$ cells follow similar trends depending on which side Li is adsorbed at. When Li is adsorbed at the top side, the binding energy is lower than for the dilute cases. This trend reverses when Li is adsorbed on the bottom side. There are two factors affecting this, which we will discuss in detail in the upcoming sections on machine learning. One is the Li-Li Coulomb and cohesive interaction and the other one is due to the dipole moment of the Janus structure which results in splitting of energies of least unoccupied state of both sides of Janus materials.

\subsection{Descriptor design for Machine Learning Model to Predict Li Adsorption Energies}

Adsorption energy is our target variable. The training set includes the adsorption energies of Li on six regular TMDs and on both top and bottom sides of six Janus TMDs. For each adsorption case, four energies are given for four different coverage ratios. This makes a total of [(6 (regular TMD) + 6 $\times$ 2 (Janus-both sides)) $\times$ 4 (coverage) ] 72 training instances.

In order to extract the key features that govern the adsorption of a metal atom on a 2D TMD structure, we decompose the energetic contributions involved in the adsorption process. Starting with an isolated Li atom and a relaxed TMD structure, adsorption at 0 K can be viewed as a process of experiencing successive potential barriers and potential wells: 

(a) First, the Li atom must overcome its ionization energy ($E_{ion}$) to transfer its charge to the vacuum.

(b) The charge is then transferred from the vacuum to the lowest unoccupied state (LUS) in the TMD structure.  $E_{LUS}$ represents the energy of this potential well.

(c) For the adsorption of multiple Li atoms to take place, each Li atom must overcome the potential barriers of Li-Li interaction, namely, the Li-Li repulsive Coulomb potential and the Li-Li cohesive potential. $E_{Li-Li}$ refers to these two interaction energies combined. 

(d) Finally, there will be an interaction between the Li atom and the TMD, which we can view in two successive steps: a distortion of the TMD as the Li atom approaches it, followed by a coupling of the positively-charged Li atom to the negatively-charged TMD. Here we generalize the finding of M. Dou and M. Fyta\cite{dou} and assume that this distortion energy can be neglected compared to the other energy terms for all TMD structures. Thus, we have only the coupling energy described though the potential energy term ($E_{coup}$) which includes all the electrostatic and quantum-mechanical interactions between the Li atom and the TMD. 

According to this analysis, the adsorption energy ($E_{d}$) can be decomposed as:

\begin{equation}
    E_d=E_{ion}+E_{LUS}+E_{Li-Li}+E_{coup}
    \label{eqn:mlall}
\end{equation}
We rely on the above energy representation to select the best descriptors for predicting the adsorption energies of Li atoms on 2D TMDs. 


\begin{figure*}
    \begin{center}
    \includegraphics[width=16 cm]{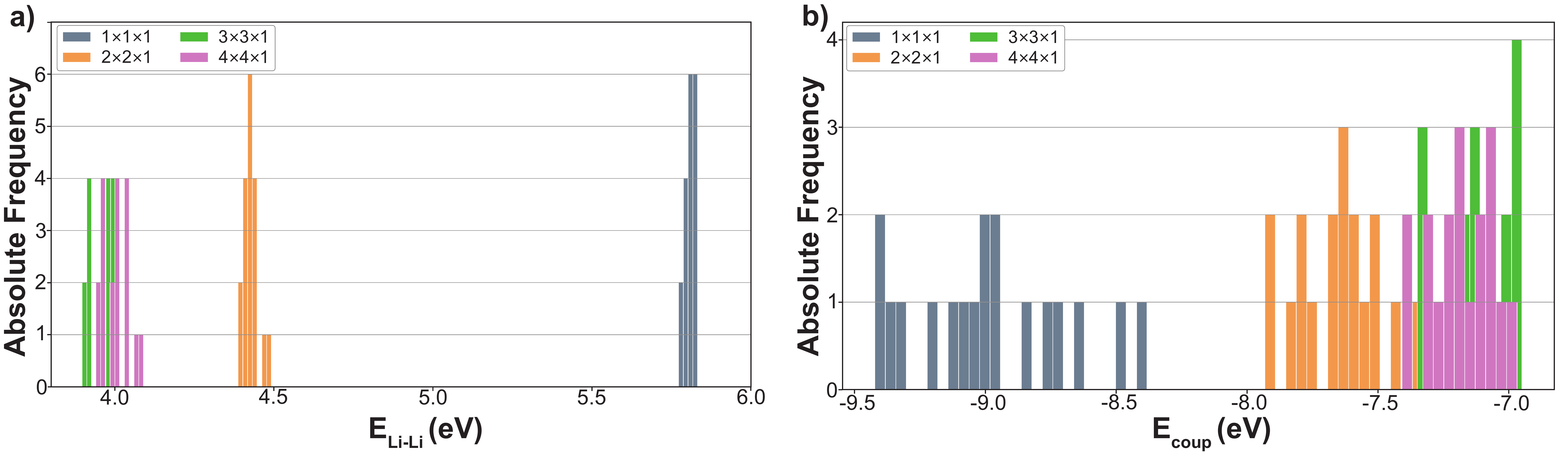}
    \caption{\label{fig:clusters}{a) Histogram of Li-Li interaction energy (including Coulomb and cohesive interactions) labeled by supercell size.  b) Histogram of Li-TMD coupling energy (the definition of this term is defined in the text).}}
\end{center}
\end{figure*}


We then try to analyze the coverage-dependent energy terms, $E_{Li-Li}$ and $E_{coup}$, in order to decouple the coverage dependence from the material dependence.  The Li-Li interaction term was calculated by conducting single atom DFT simulations of a positively charged Li$^+$ atom in a 2D cell, with 15 Å of vacuum, according to the lattice parameters of regular and Janus TMDs.  In Fig. \ref{fig:clusters}a), we notice that the distribution of $E_{Li-Li}$ forms separate clusters of continuously distributed values. These clusters correspond to the four supercell sizes we are studying. We also notice that the individual distributions of the $3 \times 3 \times 1$ and $4 \times 4 \times 1$ supercells almost entirely overlap. This reflects how a coverage-dependent energy term would behave. Hence, we claim that the same behavior would appear for $E_{coup}$. In order to avoid the direct calculation of $E_{coup}$ with DFT, we calculated all other energies on the right hand side (RHS) of Eq. \ref{eqn:mlall} and then $E_{coup}$ was calculated indirectly as the difference between the DFT-calculated adsorption energies and the energies on the RHS.  Fig. \ref{fig:clusters}b) indicates that our claim is valid and that the distribution of $E_{coup}$ can be viewed to form separate clusters of continuously distributed values similar to $E_{Li-Li}$.  We also notice that $E_{coup}$ becomes more negative (more attractive) as we increase the doping concentration due to the larger electrostatic attraction between the Li atom and the charges transferred from the Li atoms that tend to accumulate on the TMD surface. Even though the amount of charge transfer from one Li atom decreases with the doping concentration, the charge density of the negative accumulated charge on the TMD surface increases with doping resulting in a stronger coupling. This trend is opposite to that of $E_{Li-Li}$, which becomes more positive (more repulsive) with higher doping.  

Therefore, we can decouple the coverage dependence from the material dependence by categorizing our training examples according to the supercell size of the TMD structure to one of three categories ($1 \times 1 \times 1$, $2 \times 2 \times 1$, and $3 \times 3 \times 1 $ and $4 \times 4\times 1$) and then performing our ordinary least squares (OLS) regression within each category. 
Since the coverage-dependent energy terms of the $3 \times 3 \times 1 $ and $4 \times 4 \times 1$ supercells strongly overlap in Fig. \ref{fig:clusters}, it is evident that the dilute case is achieved in the $3 \times 3 \times 1 $ supercell.  Thus, we group the $3 \times 3 \times 1 $ and $4 \times 4 \times 1$ supercells into the same cluster.  Recall, that we concluded the same in our earlier discussions about bond length and charge transfer at different concentrations.

After decoupling the coverage dependence through clustering, we must still take into account the material dependence of $E_{coup}$ within each cluster which is manifested in the variance of each cluster in Fig. \ref{fig:clusters}b).  We do so by calculating the coverage-independent part of $E_{coup}$, represented by the interaction energy of a single Li atom with a bare regular/Janus TMD structure $E_{Li-TMD(bare)}$, by using a simple Coulomb interaction model. $E_{Li-TMD(bare)}$ plays an important role in displaying the intrinsic differences between regular and Janus TMDs. It is attractive when adsorption takes place on the higher-EN side of Janus TMDs, but repulsive for adsorption on the lower-EN side.  Also, $E_{Li-TMD(bare)}$ is almost zero for regular TMDs. These observations are due to the intrinsic dipole moments of Janus TMDs.  


Therefore, inside each cluster, Eq. \ref{eqn:mlall}  transforms into:
\begin{equation}
    E_d=E_{ion}+E_{LUS}+E_{Li-Li}+E_{Li-TMD(bare)}
    \label{eq:mlaccurate}
\end{equation}
For this study, we can drop off $E_{ion}$ as we are considering only Li as our adsorbate. To generalize, the predicted adsorption energy can then be written as a linear combination of the remaining energy descriptors:
\begin{equation}
E_{d}^{pred}=a E_{LUS}+b E_{Li-TMD(bare)}+c E_{Li-Li}+d 
\label{eq:mlfinal}
\end{equation}
where $E_{d}^{pred}$ is the predicted adsorption energy and a, b, c and d are parameters that are tuned through the regression process.  The energy descriptors on the RHS of this equation are calculated as follows:

(a) The energy of the lowest unoccupied state is found by subtracting the vacuum energy from the energy of the conduction band minimum  ($E_{LUS}=E_{CBM}-E_{vacuum}$) of bare TMDs.  Note that due to the intrinsic dipole moment of Janus TMDs, the vacuum level energies are different at different sides of Janus TMDs,\cite{ersan2} which explains why adsorption energies are dependent on the side of adsorption. $E_{LUS}$ varies between -4.74 and -3.18 eV depending on the Janus material and the side of Li adsorption.  This descriptor covers the effects of the electronic dipole moment on the electronic structure and is material- and adsorption side- dependent.

(b) To find the Coulomb energies of Li interaction with the bare TMDs ($E_{Li-TMD(bare)}$), we summed up the Coulomb interaction energies of a single Li atom with the atoms in the three layers of the X-M-Y structure. We assumed that Li is placed on top of three charged parallel plates (positively charged plate for M layer, negatively charged plates for X and Y layers). The charge density on the plates were determined by our Bader analysis results and the lattice constants of the monolayers. For instance, to determine $E_{Li-TMD(bare)}$ for Li adsorption on the S side of MoSTe,  we calculated and then summed up all Li-S, Li-Mo, and Li-Te layer interactions for a single Li atom over the S side of MoSTe. We also assumed that Li has a charge of +1 $e^-$. The perpendicular distance between the Li atom and the plate on which adsorption takes place was calculated by approximating the distance between the Li atom and any of the nearest neighbor chalcogen atoms as a summation of their atomic radii. According to this, Li adsorption on higher EN sides results in exothermic reactions, while adsorption on the other side results in endothermic reactions.  This descriptor is material- and adsorption side- dependent. It is expected to capture the electrostatic energy cost of the intrinsic dipoles of Janus structures.

(c) For the Li-Li interaction energies ($E_{Li-Li}$), we calculated the DFT total energy of a single +1 e$^-$ charged Li atom (+0.99 e$^-$ due to pseudopotential restrictions) arranged in a separate 2D monolayer (with 15 \AA{} of vacuum layer) making use of the relaxed lattice constants of each structure. The total energy includes Li-Li cohesion and Coulomb interactions. This descriptor is mainly concentration-dependent such that the magnitudes of this endothermic energy contribution decays significantly from dense to dilute doping concentrations.  

To check the quality of our model, we calculated the root mean squared error (RMSE), the mean absolute error (MAE), and the coefficient of determination (R$^{2}$) of our cluster-wise linear regression (CLR) model and they are tabulated in Tab. \ref{table-regression}. RMSE, MAE and R$^2$ values are also reported for linear regression without clustering in Tab. \ref{table-regression} in order to show the importance of clustering the dataset.

\begin{table}[h!]
\centering
    \caption{Linear regression statistics for each cluster. OLS-optimized values for regression parameters from Eq.\ref{eq:mlfinal}. Linear regression statistics are also tabulated for the dataset without clustering.}
    \begin{tabular}{c|c|c|c|c}
        \multicolumn{5}{c}{\textbf{Cluster-wise Linear Regression Model}}\\
       \hline
         \textbf{Supercell} & \textbf{a} & \textbf{b} & \textbf{c} & \textbf{d} \\
         \hline
         $1 \times 1 \times 1$ & 0.129  & 0.044 & -0.032 & -1.672 \\ 
         $2 \times 2 \times 1$ & 0.151 & 0.249 & 0.035 & -1.719 \\ 
         $3 \times 3 \times 1$ / $4 \times 4 \times 1$  & 0.337 & 0.084 & -0.015 & -1.663 \\ 
         \hline
         \multicolumn{5}{c}{RMSE: 0.055 eV  MAE: 0.044 eV  R$^2$: 0.977} \\
         \hline
         \multicolumn{5}{c}{}\\
        \multicolumn{5}{c}{\textbf{Linear Regression Model}}\\
        \hline
         \textbf{All Dataset} & \textbf{a} & \textbf{b} & \textbf{c} & \textbf{d} \\
         \hline
           & 0.248  & 0.103 & -0.001 & -1.679 \\ 
          \hline
          \multicolumn{5}{c}{RMSE: 0.120 eV  MAE: 0.094 eV  R$^2$: 0.876} \\
         \hline

    \end{tabular}
    \label{table-regression}
\end{table}

\subsection{Linear Regression of Adsorption Energies}


\begin{figure*}[!ht]
\includegraphics[width=16cm]{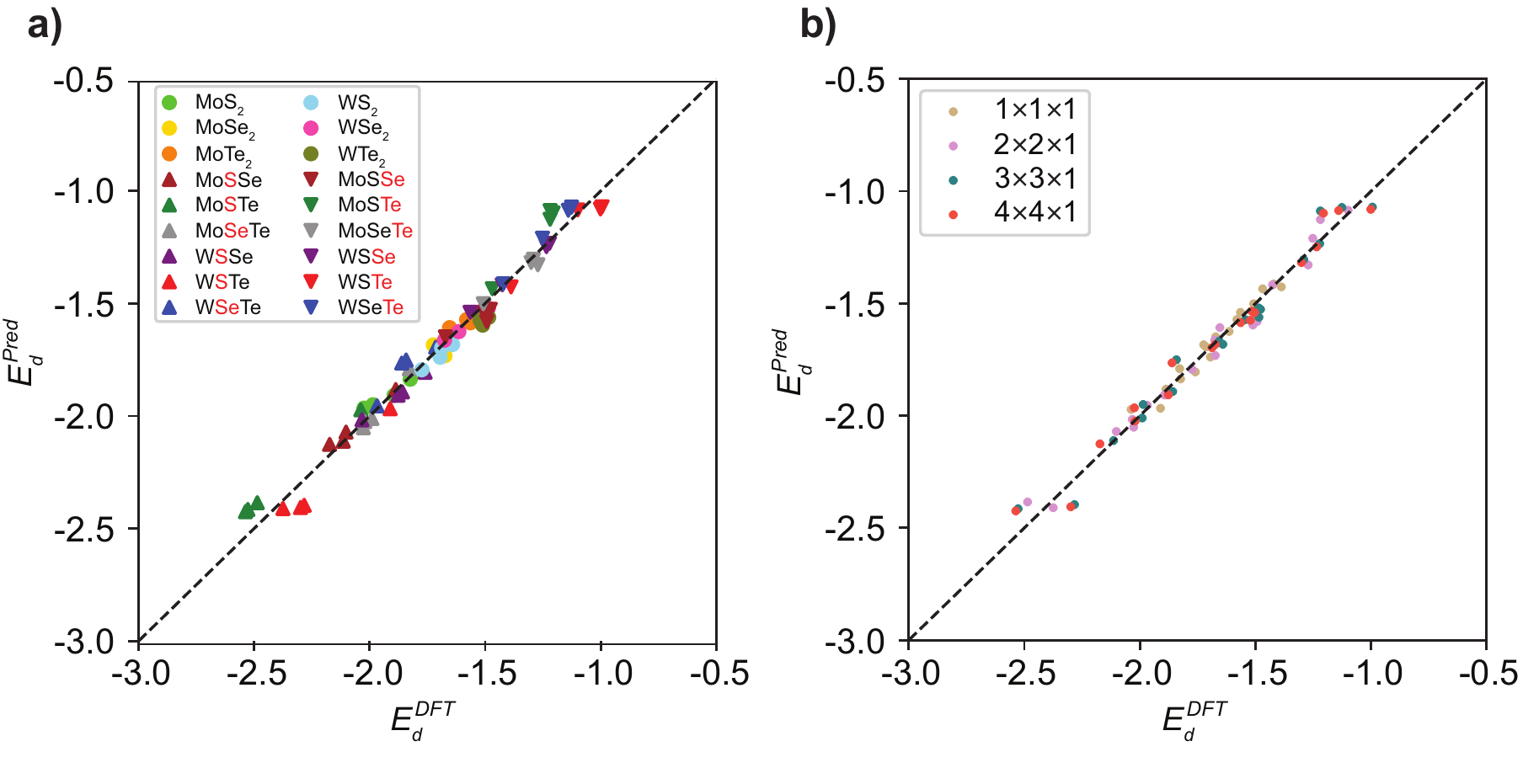}
\caption{Predictions of adsorption energies from CLR ($E_{d}^{pred}$) vs original DFT values ($E_{d}^{DFT}$). The left plot (a) classifies data according to material (color) as well as side of Li adsorption on Janus materials (up and down arrows). Data for all supercell sizes are given, but are not distinguished. The right plot (b) classifies the same data according to supercell size.}
\label{fig:ML}
\end{figure*}

For each sub-model within each cluster, we used a $3$-fold cross-validation scheme via the scikit-learn package,\cite{scikit-learn} which means that the data in each cluster was divided randomly into three sets. One set was used for testing and the other two for training.  This procedure is repeated so that every set is used once as a testing set to avoid any overfitting of the model to particular training examples. To simplify making comparisons among the different energy descriptors influencing the adsorption energy, the descriptors inside each cluster were standardized by re-scaling each descriptor to have a mean of zero and standard deviation of one. The averaged cross-validation RMSE and R$^2$ scores for the CLR model are $0.056$ eV and $0.973$, respectively. These cross-validation measures give an estimate of the generalization power of the model on future unseen data. The statistical measures for our final CLR model applied to the whole dataset are as follows: $0.055$ eV (RMSE), $0.043$ eV (MAE), and $0.977$ (R$^2$).  The predicted adsorption energies from the CLR model versus the DFT calculated energies are shown in Fig \ref{fig:ML}. We note that a direct application of linear regression without clustering would give: $0.120$ eV (RMSE), $0.094$ eV (MAE), and 0.877 (R$^2$), which proves the power of our initial clustering technique. The regression parameters of the standardized energy descriptors for the three linear sub-models inside each cluster, as well as the parameters for the non-clustered model are summarized in Tab. \ref{table-regression}.  It is important to note that the linear regression model without clustering de-emphasizes the concentration-dependent E$_{Li-Li}$ interaction term as noticed from its small $c$ value in Tab. \ref{table-regression}, which results in a significant decrease in the model accuracy.

Focusing on the CLR model, notice in Tab. \ref{table-regression} that the coefficient $a$ increases with supercell size, indicating that $E_{LUS}$ becomes a more significant descriptor as we approach the dilute case.  This finding agrees with the works by Dou \emph{et al.}\cite{dou} and Stavri{\'c} \textit{et al.} \cite{stavric2018understanding} For dilute doping concentrations, the electronic structure of the material will not be affected by doping and this descriptor can be used reliably. However, in dense doping cases, adsorbents may form extra bands around conduction and valance band regions which may result in serious deviations in the electronic structure from that of the bare TMD.  Our CLR model suppresses this effect by fitting dilute doping (giving more weight) and dense doping (giving less weight) cases separately, which makes the model physically more reliable than a linear model without clustering.

Looking at the concentration-dependent term $E_{Li-Li}$, in order to compare the significance of this descriptor to the adsorption energy in the different clusters, we need at first to take out the material dependence from the $c$ coefficients and focus only on the concentration dependence.  Recall that the regression coefficients in Tab. \ref{table-regression} are fitted to standardized descriptors. Thus, we need to divide each $c$ by the corresponding standard deviation of $E_{Li-Li}$ in each cluster. With standard deviations of $0.016$, $0.026$ and $0.045$ eV for the $1 \times 1 \times 1$, $2 \times 2 \times 1$  and $3 \times 3 \times 1 / 4 \times 4 \times 1$ clusters respectively, we note that the significance of $E_{Li-Li}$ to the adsorption energy decreases significantly as we approach the dilute doping case, as we would expect. It is also evident from the small magnitudes of $c$ that $E_{Li-Li}$ is the least important descriptor for the adsorption energy. This can be attributed to the fact that $E_{Li-Li}$ has a low material dependence as noticed from the small variance of $E_{Li-Li}$ within each cluster (see Fig. \ref{fig:clusters}a). Note also that the high dependence of $E_{Li-Li}$ on doping concentration was already decoupled through clustering.  

Based on our analysis of bond lengths (Fig. \ref{fig:distance}), charge transfer (Fig. \ref{fig:charge}) and adsorption energies (Fig. \ref{fig:ads}), we concluded that Li adsorption on the $3 \times 3 \times 1$ supercell size can be considered as dilute doping. We also noted that at dense doping ($1 \times 1 \times 1$ primitive cell) concentrations, the binding chemistry of Li was different than the dilute doping cases. Even though these properties at dilute doping concentrations ($3 \times 3 \times 1$ supercell or larger) show similarities with those of the $2 \times 2 \times 1$ supercell, the histogram of $E_{coup}$ in Fig. \ref{fig:clusters}b) shows a significant difference in $E_{coup}$ ($\sim0.6$ eV) between Li adsorption on the $2 \times 2 \times 1$ supercell and the $3 \times 3 \times 1/ 4 \times 4 \times 1$ supercell sizes.  This provides evidence that coupling of the Li adatom with the $2 \times 2 \times 1$ supercell is stronger than in the dilute cases. This is the reason why bond lengths of the Li adatom to nearest chalcogen atoms are smaller ($\sim 0.02$ Å) and the adsorption energies are larger ($\sim 70$ meV) for the $2 \times 2 \times 1$ supercell than in the dilute doping supercells sizes.  This explains why our CLR model elevates the value of the $b$ coefficient for the $2 \times 2 \times 1$ cluster, emphasizing the $E_{Li-TMD(bare)}$ term for the $2 \times 2 \times 1$ supercell above that of the other supercell sizes. Our CLR model successfully represents this intermediate doping concentration by de-emphasizing the electronic structure based descriptor ($E_{LUS}$) while increasing the importance of the concentration-dependent ($E_{Li-Li}$) and material- and adsorption side- dependent ($E_{Li-TMD(bare)}$) descriptors.

Finally, because we have standardized our descriptors, the $d$ parameter should correspond to the adsorption energy of Li on a TMD structure whose descriptors have the mean values of our dataset. The mean values of $E_{LUS}$ and $E_{Li-TMD(bare)}$ are $-3.88$ and $-0.001$ eV respectively, while the mean value of $E_{Li-Li}$ varies depending on the cluster since it is a concentration-dependent descriptor. For instance, for the $3 \times 3 \times 1 / 4 \times 4 \times 1$ cluster, $E_{Li-Li}$ has a mean value of $3.98$ eV.  These three means are very close to the descriptor values of Li adsorption on a $4 \times 4 \times 1$ supercell of MoSe$_2$. That is why we see $d = -1.663$ eV for the $3 \times 3 \times 1 / 4 \times 4 \times 1$ cluster, which is roughly the same as the adsorption energy of Li on the MoSe$_2$ TMD structure in the $4 \times 4 \times 1$ supercell (see Tab. \ref{tab:table2}).

Based on our physically interpretable descriptors, we expect our CLR model to be transferable to predict the adsorption energies of other adatoms on other 2D materials beyond regular and Janus TMDs provided that the adsorption process is dominated by a near full charge transfer from the adatom to the substrate. Recall that our model assumes that the adsorption process is governed by a simple charge transfer in which the adatom transfers its charge to the 2D material without introducing additional electronic states to the substrate. Recall also that we have fixed the Li charge to be always +1 e$^-$ in our model assuming full charge transfer. If additional electronic states are created near the lowest unoccupied state of the 2D material upon adsorption or if the charge transferred is far away from being a full charge transfer, the adsorption picture diverges from the model assumptions and the model would not be expected to give reliable predictions in such cases. Another point that needs to be satisfied to guarantee the transferability of the model is that the geometrical distortion in the structure of the 2D substrate which occurs due to the presence of the adatom has to be negligible since our model neglects this distortion energy. When the data in Fig. \ref{fig:ML} is carefully examined, it is clear that the highest difference between calculated and predicted energies arises from the structures with the highest dipole moments (MoSTe and WSTe).  Even though we included the effects of intrinsic dipole to our CLR model using electronic structure and Coulomb interaction terms, one can also enhance CLR fitting by including the structural effects such as spontaneous curvature of Janus structures in finite sizes.\cite{xiong} We came up with a simple model for Janus TMD layers to calculate the radius of curvature, taking into account the in-plane stiffness of the constituent regular TMD layers.\cite{ataca}  For example, for MoSSe, we used lattice constants and in-plane stiffness of MoS$_2$ and MoSe$_2$ to strain the combined structure to curve and match the lattice constants of MoS$_2$, MoSSe and MoSe$_2$ on S, Mo and Se layers of the Janus structure. Once the radius was determined, we calculated the energy cost in terms of strain and placing Li adatom inside and outside the curved structure (Coulomb interaction). After including this in our CLR model, we managed to reduce RMSE to 0.045 eV,  MAE to 0.037 eV and increase $R^2$ to 0.984. When compared with our reported values in Tab. \ref{table-regression}, these are minor improvements to the CLR model. We do not focus on this method further because: i) This method is Janus structure specific and the effects of curvature are calculated primitively instead of giving a full quantum mechanical treatment. ii) Our aim is to develop a ML descriptor representation which takes into account the effects of coverage and different alloys of the underlying 2D layer. Calculating the in-plane stiffness in accordance with the alloying concentration is a process that would be extremely computationally demanding. This conflicts with the goal of using ML to make accurate predictions without the high computational cost of quantum mechanical simulations.

\subsection{Suitability of Janus TMDs for next generation anode material}

In this section, we will investigate if Janus TMDs are good candidates for next generation anode materials. In order to achieve conclusive results, we will focus on three different criteria including the ease of Li diffusion on the Janus TMD surfaces, prediction of open circuit voltage (OCV), storage capacity and effects of multilayer Janus systems, and  volumetric expansion during dis/charge cycle and the effects of Li on the electronic structure.

\subsubsection{i) Li Diffusion on Monolayers}

\begin{figure*}
\begin{center}
\includegraphics[width=15cm]{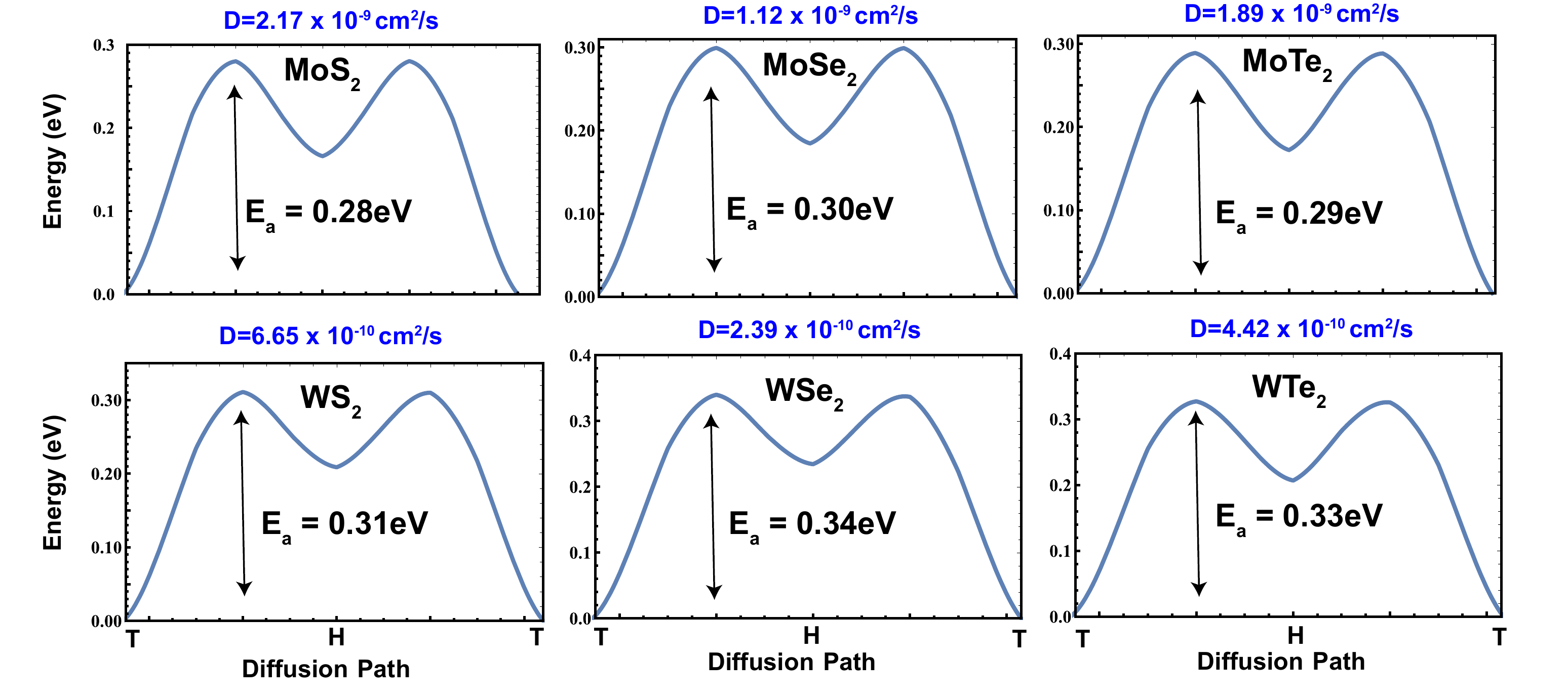}
\caption{\label{fig:neb-bare} Activation energy barriers for Li atom diffusion for regular TMD-MX$_2$ (M =Mo, W; X = S, Se, Te),  calculated with NEB simulations. The relative energies with respect to the ground state adsorption geometry are given along y axis. Diffusion path is from top of metal site through hollow site to neighboring top site as indicated in x-axis. See Fig.\ref{fig1-adsorption} for adsorption geometries.}
\end{center}
\end{figure*}

\begin{figure}
\begin{center}
\includegraphics[width=11cm]{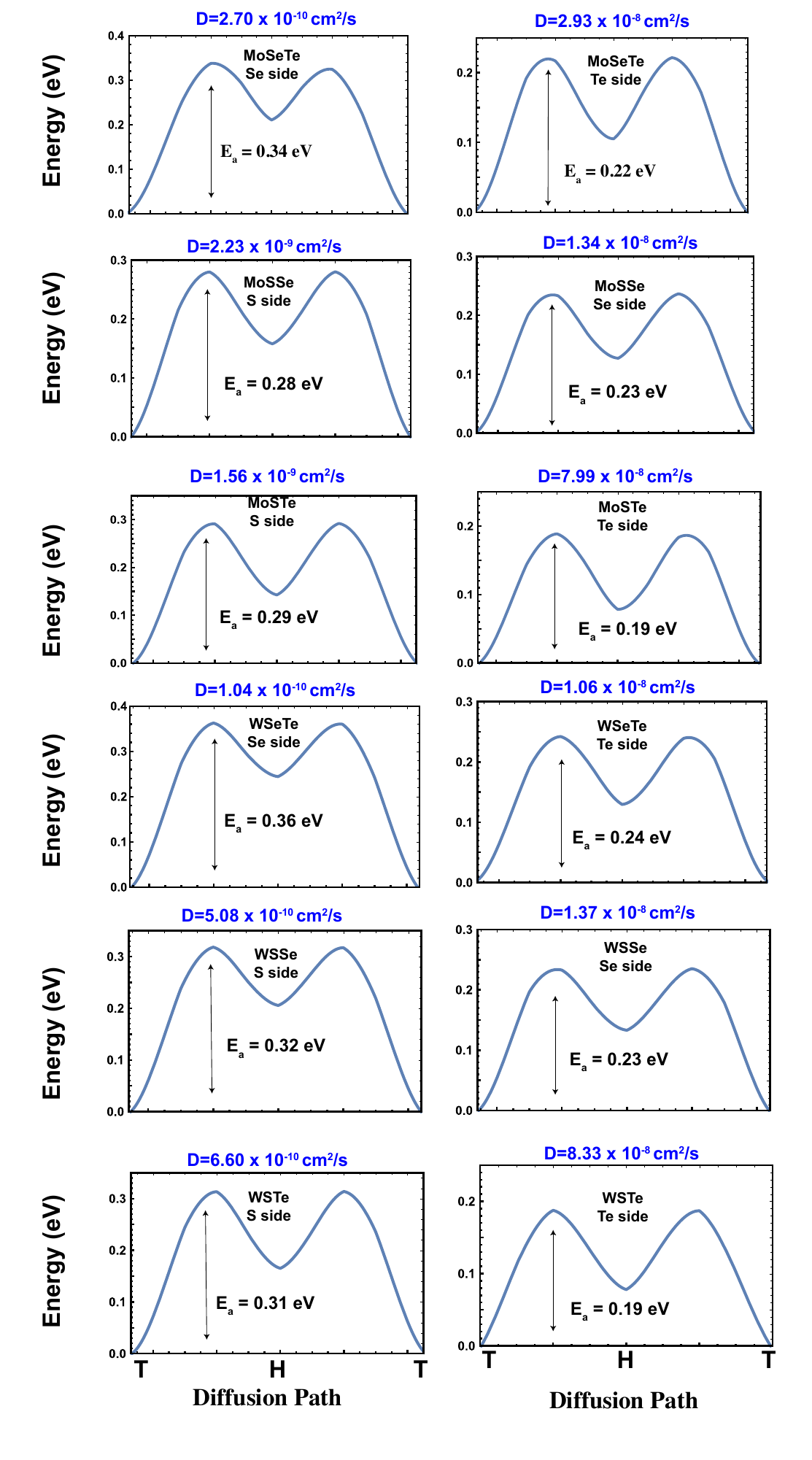}
\caption{\label{fig:neb-janus}Activation energy barriers for Li atom diffusion for Janus TMDs MXY (M = Mo, W; X/Y = S, Se, Te), calculated with NEB simulations.  Barriers for top and bottom sides are calculated separately.  The relative energies with respect to the ground state adsorption geometry are given along y axis. Diffusion path is from top of metal site through hollow site to neighboring top site as indicated in x-axis. See Fig.\ref{fig1-adsorption} for adsorption geometries.}
\end{center}
\end{figure}

In addition to adsorption, Li migration on the monolayers determines the structures' abilities as anodes in Li ion batteries.  The climbing image nudged elastic band method (NEB) simulations for Li on regular TMD structures and Janus materials were given in Fig. \ref{fig:neb-bare} and Fig. \ref{fig:neb-janus}, respectively. In both figures, only the $3 \times 3 \times 1$ supercell case is given.  This is justifiable because the adsorption energies converge around $3 \times 3 \times 1$ supercell sizes, as discussed in the previous sections.  For all of the examined configurations, the adsorption energy is maximum for Li atoms above the metal top sites. The second most favorable site is the hollow site. This implies that Li atoms transverse the monolayer by hopping between metal top sites while passing over metastable hollow sites. This is true for every supercell size and monolayer side. From the NEB plots, we find the activation energy barrier that a Li atom must pass over to migrate from a top-metal site (at the zero point) to a hollow site (the middle valley). These values are recorded in Tab. \ref{tab:neb}, along with the Li adsorption energies at the top-metal site, the diffusion coefficients, and the charge lost by the Li after adsorption.  From the energy barriers, we obtained the diffusion coefficients, which is the frequency with which Li ions move between nearest neighbor top-metal sites.  It is defined as  

\begin{equation}
    D = a^{2}ve^{-E_a/{k_{B}T}}
 \label{eq:diffusion}    
\end{equation}
where \textit{a} is the lattice constant, \textit{v} = \num{1e11} Hz is the order of vibration of the Li adatom, \textit{$E_a$} is the activation energy barrier that must be overcome for diffusion to occur, \textit{$k_B$} is the Boltzmann constant, and \textit{T} is temperature (300 K).\cite{ersan}  The product in front ($a^{2}v$) acts as the probability of a successful jump between sites.\cite{toyoura}  As the energy barrier decreases or lattice constant increases, the diffusion coefficient increases.  Thus anode materials with low diffusion energy barriers experience higher Li ion mobility, which is expressed as higher diffusion coefficients. Using Eq. \ref{eq:diffusion}, we calculated the diffusion coefficient for graphene as \num{8.72e-09} cm$^2$/s, where we used a lattice constant of 2.465 \si{\angstrom} from Leggesse \textit{et al.}’s DFT study\cite{leggesse} and the energy barrier from Zhong \textit{et al.}'s study.\cite{zhong}  In addition, the diffusion coefficients we calculated for six TMDs range from \num{2.39e-10} cm$^{2}$/s for WSe$_2$ to \num{2.17e-09} cm$^{2}$/s for MoS$_2$.  This implies that MoS$_2$, MoSe$_2$, and MoTe$_2$ have diffusion coefficients that have the same order of magnitude to that of graphene.  Thus, Li ions on these TMDs have about the same mobility as on graphene. 

\begin{table}[h!]
\centering
    \caption{Li on $3 \times 3 \times 1$ supercell Janus Monolayers; Side where lithiation occurs, adsorption energy (E$_d$), activation energy (E$_a$), diffusion coefficient (D), change of charge on Li atom after adsorption ($\Delta \rho_{Li}$).  The adsorption energies are taken from Tab. \ref{tab:table2}}
    \begin{tabular}{c|c|c|c|c|c}
        \textbf{TMD} & \textbf{Side} & \textbf{E$_d$(eV)} & \textbf{E$_a$(eV)} & \textbf{D(cm$^2$/s)} & \boldmath$\Delta \rho_{Li} (e)$\\
         \hline
         MoS$_2$ & -  & -1.986 & 0.28 & \num{2.17e-9}& 0.874 \\ 
         MoSe$_2$ & - & -1.657 & 0.30 & \num{1.12e-9}& 0.866 \\
         MoTe$_2$ & -  & -1.544 & 0.29 & \num{1.89e-9}& 0.860 \\
         WS$_2$ & -  & -1.641 & 0.31 & \num{6.65e-10}& 0.868 \\
         WSe$_2$ & -  & -1.486 & 0.34 & \num{2.39e-10}& 0.860 \\
         WTe$_2$ & -  & -1.485 & 0.33 & \num{4.42e-10} & 0.859 \\
         
         MoSeTe & Se & -1.990 & 0.34 & \num{2.70E-10} & 0.862\\
         MoSeTe & Te & -1.291 & 0.22 & \num{2.93E-8} & 0.865\\
         MoSSe & S & -2.114 & 0.28 & \num{2.23E-9} & 0.870\\
         MoSSe & Se & -1.478 & 0.23 & \num{1.34E-08} & 0.870\\
         MoSTe & S & -2.527 & 0.29 & \num{1.56E-09} & 0.868\\
         MoSTe & Te & -1.219 & 0.19 & \num{7.99E-08} & 0.866\\
         
         WSeTe & Se & -1.842 & 0.36 & \num{1.04E-10} & 0.857\\
         WSeTe & Te & -1.126 & 0.24 & \num{1.06E-08} & 0.859\\
         WSSe & S & -1.858 & 0.32 & \num{5.08E-10} & 0.867\\
         WSSe & Se & -1.222 & 0.23 & \num{1.37E-08} & 0.859\\
         WSTe & S & -2.283 & 0.31 & \num{6.60E-10} & 0.867\\
         WSTe & Te & -0.994 & 0.19 & \num{8.33E-08} & 0.857\\
    \end{tabular}
    \label{tab:neb}
\end{table}

Notice that the Y-side MXY cases feature higher diffusion coefficients than their parent TMDs. Even though the lattice constants of parent TMDs are larger, the diffusion barriers on Y-side of Janus MXYs are always lower which result in enhanced diffusability of Li on the surface.  For instance, Li on the Te side of MoSeTe experience a larger diffusion coefficient (\num{2.93E-08} cm$^{2}$/s) than on MoSe$_{2}$ (\num{1.12e-9} cm$^{2}$/s) and MoTe$_{2}$ (\num{1.89e-9} cm$^{2}$/s). The Janus monolayers' relatively high diffusion coefficients are mostly due to their small activation barriers.  Li across the Te side of WSTe encounters the smallest energy barrier (0.19 eV) and the highest diffusion coefficient (\num{8.33E-08} cm$^{2}$/s) of all the examined materials. 

In the case of Li diffusion on the X-side of the Janus structures, diffusion coefficients are either very alike (WSSe and WSTe) or smaller compared to their parent TMDs. (except S side of MoSSe which has a slightly higher diffusion coefficient than that of MoS$_2$) This is due to the fact that all of the calculated activation energies, $E_a$, on the X-Side of the Janus structures are higher than that of the parent TMDs. This is a consequence of having higher  adsorption energies of Li adatoms on the X-side of Janus structures. Even though the lattice constants are increasing (as discussed in  previous subsections), the effect of this on the diffusion coefficients is not pronounced. In summary, our simulations showed that the Janus monolayers behaves differently depending on where the Li adatom is adsorbed. The Y sides of the Janus TMDs are preferable to the X-sides, as Li binding energies and activation barriers are lower on the Y sides. Sulfur based surfaces on the X-side of Janus TMDs have comparable diffusion coefficients to that of their parent TMDs.



\subsubsection{ii) Voltage Profiles and Storage Capacities}

So far, our focus has been on adsorption energies and understanding the binding chemistry of single Li adsorption on Janus surfaces. However our adsorption energy definition doesn't take into account the stability and crystal structure of the adsorbant. On the other hand, the formation energy is the energy required to dissociate a system into its component parts.  Thus, stable systems have negative formation energies.  For our study, this term is defined as the energy required for an Li atom to dissociate from its bulk form and to be absorbed on the layered materials' surfaces.  We calculated the formation energy as:
\begin{equation} \label{formationEnergy}
    E_{f} = E_{\mathrm{MXY} + \mathrm{Li}} - E_{\mathrm{Bulk-Li}}/2 - E_{\mathrm{MXY}}
\end{equation}
where $E_{\mathrm{MXY} + \mathrm{Li}}$ and $E_{\mathrm{MXY}}$ are the total energies of the layered system with and without Li adsorption, and $E_{\mathrm{Bulk-Li}}$ is the energy of the lowest energy bulk structure of Li, which consists of two atoms per unit cell. The calculated formation energies are given in Table \ref{tab:table2}. According to Eq. \eqref{formationEnergy}, if the system has negative formation energy, Li adatoms can dissociate from the bulk Li structure and bind to the monolayer surface.  Positive values indicate that Li prefers to form clusters on the surface and grow into a bulk system, rather than just binding to the surface. Based on this stability analysis, MoS$_2$ is the only bare TMD which can be suitable for battery applications at varying Li concentrations. Unlike the bare TMD structures, all of the X-sides of the Janus MXY structures resulted in negative formation energies, which opens material possibilities for battery anodes. 

To further investigate this claim, we focus on calculating voltage profiles (specifically open circuit voltage-OCV) upon the charging and discharging processes of the candidates. OCV values that are positive throughout the charging/discharging processes indicate energetic stability of adsorption on Janus TMD surfaces. In order to calculate this quantity accurately, one should have concentration dependent ground state energetics. For this reason, we performed a cluster expansion\cite{CE_Zunger, CE_vandeWalle} simulations to find the lowest energy distribution of Li atoms on the X surfaces of MXY supercells, as a function of concentration. We did not study the Y surfaces, as our formation energy data suggest that Li prefers clustering at Y surfaces. Since we focus on adsorption of Li on free-standing monolayers, the formation energies may vary when these structures are on substrates which might make the Y-side favorable for Li adsorption.

\begin{figure*}
    \begin{center}
    \includegraphics[width=15cm]{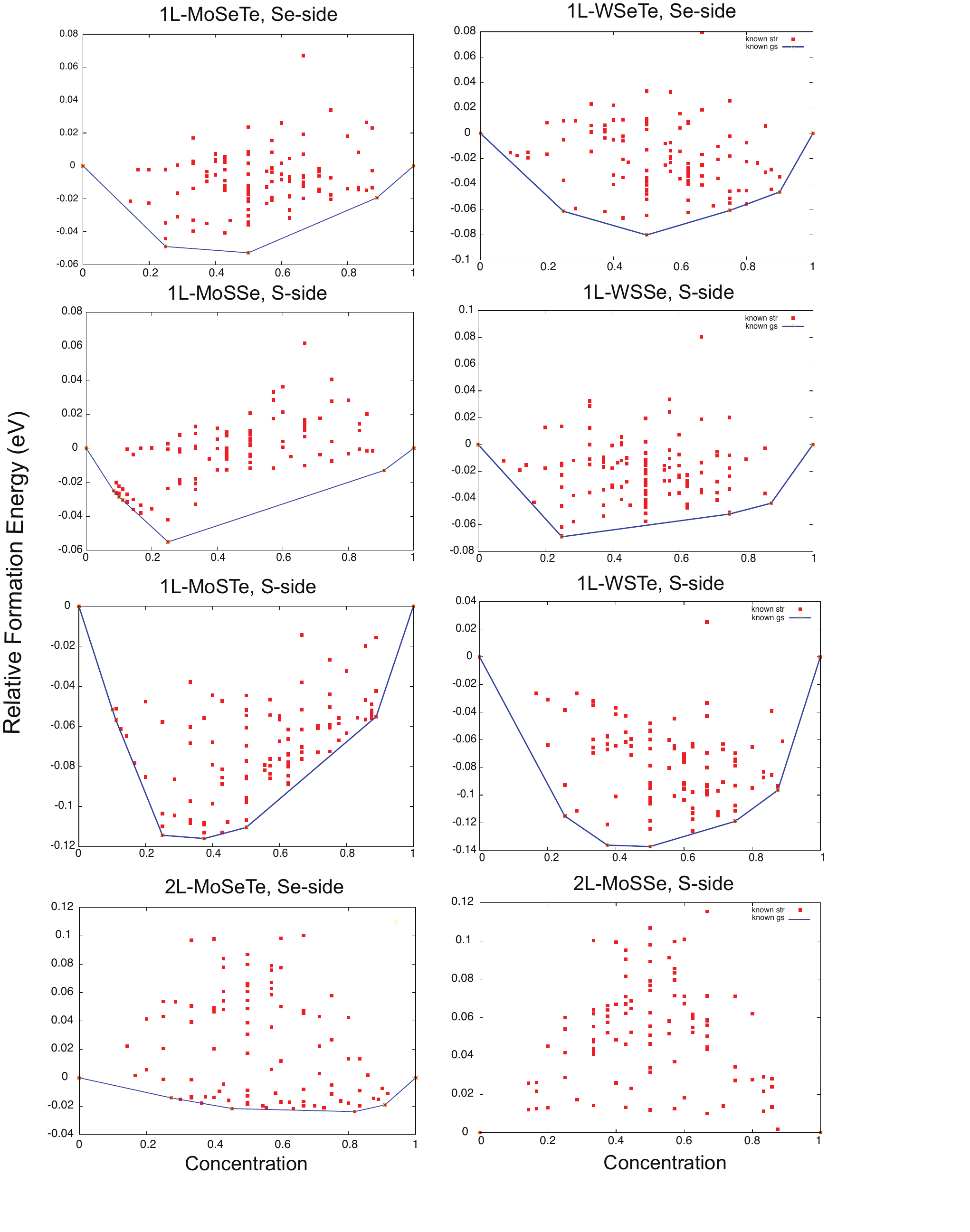}
    \caption{\label{fig_CE}{Calculated relative formation energies per formula unit as a function of concentration of Li on X-sides of MXYs. 1L and 2L in the naming scheme indicate the number of layers (single-1L or double-2L) of Li adatoms on the Janus surface. X-axis represents the Li concentration on the indicated layer. Up to two layers is shown for the Se-side of MoSeTe and S-side of MoSSe.  Every red square indicates a simulated structure. The convex hull is given by the blue curve, and lies on the lowest energy states.}}
\end{center}
\end{figure*}

Figure \ref{fig_CE} denotes the calculated relative formation energies (per formula unit) as a function of Li concentration obtained from the cluster expansion calculations using the ATAT code.\cite{avdw:atat2,avdw:atat,avdw:maps}  The relative formation energy per formula unit, $E_{rf}$, is calculated as:
\begin{equation}
    E_{rf} = E_{MXY+xLi}- (1-x) E_{MXY}- x E_{MXY+Li}
 \label{eq:CE}    
\end{equation}
where $x$ is the coverage concentration on the outer Li layer, $E_{MXY+xLi}$ is the total energy of the system per formula unit, $E_{MXY}$ is the total energy of the system per formula unit in which indicated Li layer has not adsorbed Li ($x=0$) and $E_{MXY+Li}$ is the total energy of the system per formula unit in which indicated Li layer full coverage (1 Li adsorption for every MXY, $x=1$). In this figure, the convex hull connects the lowest energy structures that are the most likely to form in experiments. In other words, the convex hull identifies thermodynamically stable and homogeneous structures at T = 0 K. We considered at least 100 structures generated by Special Quasirandom Structures (SQS) method\cite{CE_Zunger, CE_vandeWalle} to search for energetically favorable structures for each considered system. The cross validation errors, measuring the predictive power of cluster expansion, are as small as 5 meV per cell, implying an accurate prediction of the convex-hull. All of our CE results, on one layer of Li adsorption and on the X-side of Janus structures, are reported on the first 3 rows of Fig.\ref{fig_CE} and feature stable intermediate structures, which indicate homogeneous adsorption or disorption of Li atoms on the surface during charging and discharging processes.


\begin{figure*}
    \begin{center}
    \includegraphics[width=17cm]{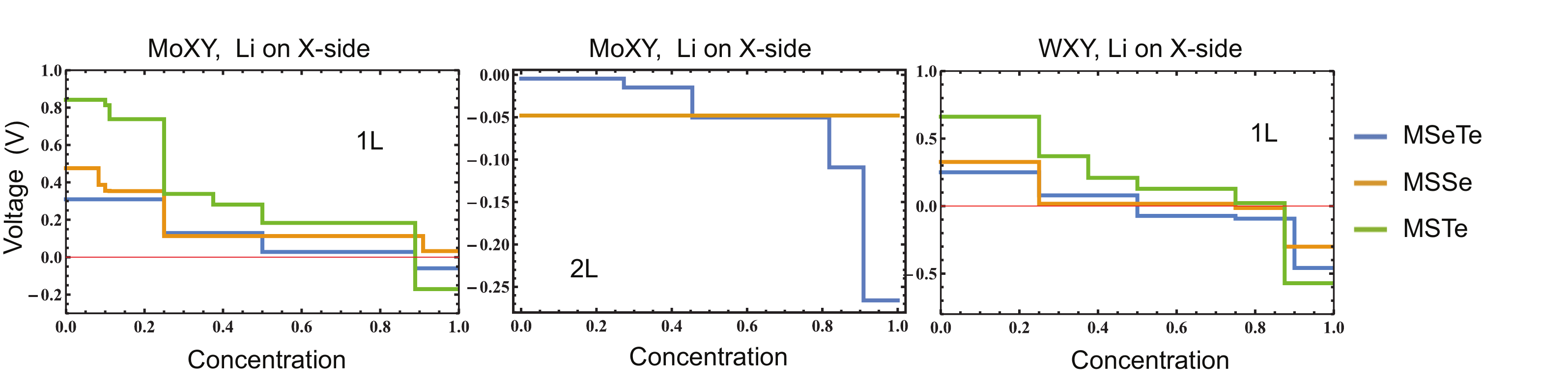}
    \caption{Voltage profiles as a function of Li concentration, for Li on the X-sides of MXYs.  1L and 2L represent one and two layers of Li coverage on the Janus surface. Concentration represents the number of Li atoms per formula unit for the indicated layer. Blue, yellow and green lines correspond to voltage profiles of MSeTe, MSSe and MSTe, respectively. Red line indicates the 0 V at every concentration.}
    \label{fig_voltage}
\end{center}
\end{figure*}

Defined as the chemical potential difference between the cathode and anode, cell output voltage is one important indication of battery performance. In this work, we investigated the half cell reaction,
\begin{equation}
    \mathrm{Li}_{x_1}\mathrm{MXY} + (x_1-x_2)\mathrm{Li} \xrightarrow{} \mathrm{Li}_{x_2}\mathrm{MXY}
\end{equation}
where $x_1$ and $x_2$ are the numbers of adsorbed Li (per formula unit) before and after the reaction.  The average anode voltage is thus computed as:

\begin{equation}
    \Bar{\mathrm{V}} = -\frac{E(\mathrm{Li}_{x_2}\mathrm{MXY}) - E(\mathrm{Li}_{x_1}\mathrm{MXY}) - (x_2-x_1)E(\mathrm{Li})}{(x_2-x_1)e}
\end{equation}

where $x_2 > x_1$, $E(\mathrm{Li}_{x_1}\mathrm{MXY})$ and $E(\mathrm{Li}_{x_2}\mathrm{MXY})$ are the total energies per formula unit of the anode before and after the reaction in eV, $E(\mathrm{Li})$ is the cohesive energy of a single Li atom in eV, and $e$ is the unit electronic charge.\cite{voltage_calc}  In the calculation of open circuit voltages, we considered thermodynamically stable compounds formed on the convex-hull obtained from the cluster expansion calculations.  A positive $\Bar{\mathrm{V}}$ implies sustainable charging/discharging process of the anode. Negative voltages indicate that the discharging product is less stable than the current state of the anode, resulting in an endothermic reaction to discharge more.\cite{voltage} The voltage profiles for the considered systems of single layer of Li adsorption are given in the first and third columns of Fig. \ref{fig_voltage} as a function of coverage.  Around 90$\%$ coverage, most of the structures acquire a negative voltage.  WSeTe acquires a negative voltage after 50$\%$ coverage.  Only MoSSe maintains a small, positive voltage after a full layer of Li on its S-side. It is important to note that the amplitude of OCV of on Se side of MoSeTe structure after 88 $\%$ is $\sim -50$ mV. This value (per Li) is very close to thermal energy fluctuations at room temperature. ($\sim 26$ meV) Thus, both cluster expansion and voltage calculations suggest that MoSSe and MoSeTe may be the only Janus structures that can possibly maintain multiple layers Li atoms, and can be of use as Li-ion battery electrodes. 

Furthermore, we conducted multilayer Li coverage simulations on the X-sides of MoSSe and MoSeTe. Before running CE and OCV simulations, we calculated the formation energy of Li atoms on the second layer for dilute (single Li atom per $3 \times 3 \times 1$ supercell) and dense (single Li atom per primitive cell) concentrations. The calculated formation energies range from -0.069/-0.032 eV (dense concentration) to -0.152/-0.093 eV (dilute concentration) for MoSSe and MoSeTe, respectively. Motivated by these results, we conducted CE simulations at the second Li layer in order to find the ground state adsorption geometries at varying concentrations. The bottom row of Figure \ref{fig_CE} indicates that the convex hull for energetically favorable structures only occurs on MoSeTe monolayers. It is notable that lowest energy structures at varying concentrations of MoSSe structures lie below 10 meV/formula unit. Statistically, these structures might occur at room temperature. The middle column of Fig. \ref{fig_voltage} shows the calculated OCV of Li adsorption on the second Li layer. Even though the reported values are all negative, they are below 50 mV (concentrations below 82 $\%$ for MoSeTe and 100 $\%$ for MoSeTe). Since this value (per Li) is comparable to room temperature thermal energy fluctuations, statistically there is a chance of observing these structures at ambient conditions.

\begin{figure*}
    \begin{center}
    \includegraphics[width=17cm]{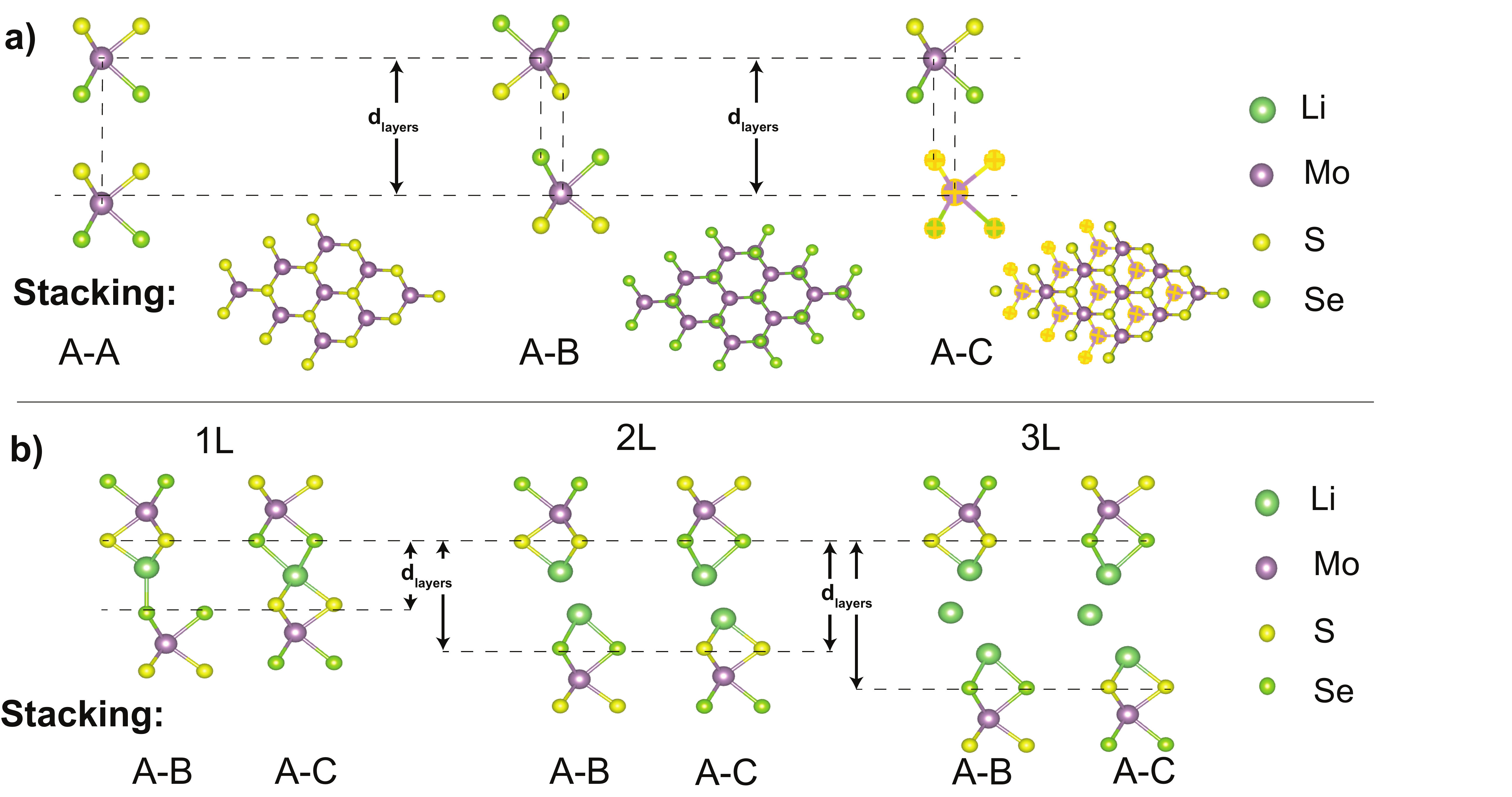}
    \caption{Stacking orientations of bilayer structures shown on SeMoS-SeMoS atomic layer configuration. a) represents pristine bilayer structures and b) represents the atomic orientations of A-B and A-C stacking after Li adsorption. 1l, 2L and 3L represent one, two and three full layers of Li adsorption between layers of bilayer MoSSe. $d_{layers}$ represents the distance between the chalcogen layers facing each other.}
    \label{fig_stacked}
\end{center}
\end{figure*}

Next, we examined how the multilayer stacking of Janus structures affects the Li storage capacity. There can be many different configurations of heterostructures of Janus TMDs, but for simplicity, we only consider bilayer configurations of MoSSe structures because it is the only material that can have full coverage on the X-side in the free standing form. Before proceeding further, we first investigated the possible bilayer orientations. We calculated the energetics of bilayer configurations where S atoms of individual layers are facing each other (SeMoS-SMoSe) and S atoms of one layer are facing the Se layer of the other layer (SeMoS-SeMoS). In each bilayer configuration, we tried different orientations of the layers: A-A, A-B, and A-C.   A-A stacking is where the same types of atoms on different layers (X/Y and M-site) are on top of each other.  A-B is where different types of atoms on different layers are on top of each other.  A-C is where the M-site of the top layer sits on top of H-site of the bottom layer, while X/Y atoms of top layer sit over the M-atom of the bottom layer. Figure \ref{fig_stacked} a) illustrates these orientations on a SeMoS-SeMoS bilayer configuration.

\begin{table*}[h!]
\centering
    \caption{Energetics of Li adsorption between the layers of bilayer MoSSe structure. A-A, A-B and A-C stacking orientations together with Li adsorption are detailed in the text and Fig.\ref{fig_stacked}. $E_b$ is the layer binding energy in eV, $d_{layers}$ is the distance between chalcogen layers facing each other in the bilayer in \AA{}, $E_f$ is the formation energy of Li adsorption per Li per formula unit in eV, and OCV is the calculated open circuit voltage in V. See the text for an explanation of the OCV calculations.}
    \begin{tabular}{c|c|c|c|c|c|c}
          & \textbf{Stacking} & \textbf{$\#$ of Li Layers} & \textbf{E$_b$ (eV)} & \textbf{d$_{layers}$ (\AA{})} & \textbf{E$_f$ (eV)} & \textbf{OCV (V)}\\
         \cline{2-7}
         
               \multirow{9}{*}{\rotatebox[origin=c]{90}{ \textbf{S}MoSe-\textbf{S}MoSe}}
        & A-A & 0 & -0.134  & 3.66 & - & -  \\  \cline{2-7}
         
        &  \multirow{4}{*}{A-B} & 0 & -0.207  & 3.11  &   -    &   -   \\
                              & & 1 &    -    & 4.08  & -0.541 & 0.541 \\
                              & & 2 &    -    & 6.04  & -0.306 & 0.072 \\
                              & & 3 &    -    & 8.39  & -0.176 & -0.084 \\
                               \cline{2-7} 
        & \multirow{4}{*}{A-C} & 0 & -0.209  & 3.08  &   -    &   -   \\
                             & & 1 &    -    & 3.60  & -0.734 & 0.734 \\
                             & & 2 &    -    & 6.32  & -0.304 & -0.126 \\
                             & & 3 &    -    & 8.33  & -0.179 & -0.072 \\
                             \hline\hline
         
                        \multirow{9}{*}{\rotatebox[origin=c]{90}{ SeMo\textbf{S}-\textbf{S}MoSe}}
        & A-A & 0 & -0.119  & 3.66 & - & -  \\  \cline{2-7}
         
        &  \multirow{4}{*}{A-B} & 0 & -0.178  & 3.09  &   -    &   -   \\
                              & & 1 &    -    & 3.92  & -0.661 & 0.661 \\
                              & & 2 &    -    & 5.88  & -0.407 & 0.153 \\
                              & & 3 &    -    & 8.33  & -0.243 & -0.086 \\
                               \cline{2-7} 
        & \multirow{4}{*}{A-C} & 0 & -0.180  & 2.98  &   -    &   -   \\
                             & & 1 &    -    & 3.41  & -0.890 & 0.890 \\
                             & & 2 &    -    & 5.86  & -0.402 & -0.086 \\
                             & & 3 &    -    & 8.17  & -0.246 & -0.065 \\
                              \hline\hline
    \end{tabular}
    \label{tab:bilayer}
\end{table*}

The layer-layer binding energy, $E_b$, is calculated as the total energy difference between the bilayer and double the energy of the monolayer Janus structures. Table \ref{tab:bilayer} reports the layer binding energy of pristine structures, $E_b$, distance between inner chalcogen layers of bilayer, $d_{layers}$, formation of energy of Li adsorption, $E_f$, and OCV. Comparing the stacking orientations, in both bilayer configurations, A-A stacking has the least binding energy and the highest layer-layer distance. The binding energies and layer-layer distances of A-B and A-C orientations are within 2 meV and 0.1 \AA{} of each other which indicates that both stacking orientations can be observed at ambient conditions. For this reason, we studied Li adsorption in both stacking orientations. It is important to note that SMoSe-SMoSe bilayer configuration is always energetically more favorable than SeMoS-SMoSe bilayer configuration. These findings are similar to what has been reported in literature.\cite{shang} Due to the method experimentalists used for synthesizing these materials from bare TMDs\cite{zhang,lu}, we believe that both bilayer configurations can be synthesized. For this reason we will focus our attention to both of the bilayer configurations.

Independent of stacking orientations and bilayer configurations, the formation energies of an added Li layer between the MXY layers are all negative, which means that clustering of Li atoms in between the MXY layers is prohibited. However, depending on the stacking orientation (ie. A-C stacking), adding a second Li layer results in negative OCV values, which influences the cyclic performance and stability of the material. Adding the 3rd layer of Li decreases OCV even more, so that such an anode is unable to discharge.  

We calculated the theoretical specific capacity of the Janus structures by using the following relation:
\begin{equation}
    C=\frac{n_i N_A n_e e}{m}
\end{equation}
where $n_i$ is the number of Li ions in the anode, $N_A$ is the Avogadro's constant, $n_e$ is the valance of ions, $e$ is the electric charge of an electron and $m$ is the weight of the anode with Li intercalation. Calculated specific capacities for monolayer and free-standing Janus structures are  125.3, 66.6, 90.74, 66.92, 76.99 and 33.73 mAh/g for MoSSe, WSSe, MoSTe, WSTe, MoSeTe and WSeTe, respectively. It is important to note that our formation energy analysis resulted in Li clustering on the Y-side of the Janus structure, which prohibits the material to be used as anode. Also, by using CE simulations, we manage to monitor the OCV during gradual charging/discharging rates. These resulted that full coverage of single layer of Li on some of the Janus structure even can't be achieved due to negative OCV. Based on these results, we conducted second layer of Li adsorptions simulations only for MoSSe and MoSeTe structures, but the calculated OCVs are all negative. We found that an A-B stacking of a MoSSe bilayer can adsorb up to two layers of Li in between the layers. Taking into consideration all of these results, we calculated the maximum theoretical specific capacity of 184.9 mAh/g for a bilayer of MoSSe in the SMoSe-SMoSe configuration and A-B stacking, with two intercalated layers of Li between the MXY-layers and an additional Li layer on the outer S side of the bilayer. 

C. Shang \textit{et al.}\cite{shang} studied single and double layer MoSSe Janus structures for Li adsorption theoretically and reported that storage capacities for single- and double-layer MoSSe can reach up to 776.5 and 452.9 mAh/g which are considerably higher than our reported values. There are two reasons for the discrepancy of the results. Firstly, C. Shang \textit{et al.} focused on the average formation energy per Li on both sides of Janus MoSSe. For that reason, even though the formation energies of Li adsorption on Y side of Janus structures are positive (see Tab.\ref{tab:table2}), the average formation energy is still negative due to the strong interaction of the X-side with Li. This enabled Shang \textit{et al.} to adsorb multiple layers of Li to the Y-side of the Janus structure as well.  The other difference is the way we calculate the OCV. We used the CE method to predict the accurate atomic ordering and energetics at varying concentrations.  Other researchers focused only on the dense concentrations at each adsorbed Li layer. Their OCV calculations only provide data on adding/removing layers of Li, but not partial Li removal inside each layer. For this reason we believe that our method provides a higher resolution and accuracy on the charging/discharging process. However, our method can further be improved by running CE simulations of multiple layers of Li not only layer by layer, but also simultaneously. This will enable modeling of partially filled multiple Li layers on the Janus surface and can avoid the high Coulomb repulsion at dense doping concentrations of each Li layer. Another method of enhancing the storage capacity of Janus structures of battery applications is to heterostructure them with other layered materials. For example Lin \textit{et al.}\cite{lin} reported that Janus MoSSe and graphene heterostructures can enhance the Li storage capacity to 560.59 mAh/g. Another study by Zhang \textit{et al.}\cite{zhang} indicated that Janus SnSSe and graphene heterostructures could achieve Li storage capacities up to 472.66 mAh/g.

\subsubsection{iii) Volumetric and Electronic Stability}

Many Li-ion batteries struggle with anomalous volume expansion, in which anode materials expand and fracture upon reaction with Li.  Such fracturing is devastating to a battery's performance, as it impedes the movement of Li ions across the electrodes and thus causes the battery's capacity to fade.\cite{lee, tahmasebi1, zhangD}  Table \ref{tab:table1} tabulates the lateral lattice constants of Janus structures, which do not change upon adsorption of a single Li atom. The thickness of Janus structures varies between 3.22 to 3.48 \AA{}. Adsorbing a single layer of Li on monolayers results in a thickness increase up to 1.74 \AA{} ($\sim 1.55 $ \AA{} for MSTe, $\sim 1.65 $ \AA{} for MSSe $\sim 1.74 $ \AA{} for MSeTe) Taking into account our two layers of Li adsorption on MoSSe and MoSeTe, the second layer of Li adds an additional $\sim 2.1 $ \AA{}. Similar observations are also carried out for bilayer MoSSe as well.  Layer-layer distances at various bilayer configurations and stacking orientations are reported in Tab. \ref{tab:bilayer}. It is important to note that adding a single layer of Li between bilayers does not result in a thickness increase ($\sim 0.9 $ \AA{} for A-B stacking and $\sim 0.5 $ \AA{} for A-C stacking) as in the case of freestanding monolayers. Adding an additional layer of Li increases the thickness by $\sim 2 $ \AA{}. These observations conclude that in order to avoid significant volumetric change during charging/discharging, bilayer Janus structures are more beneficial. As noted in the previous sections, only single layer of Li adsorption is stable on most of these structures, which suggests less volumetric change during charging/discharging process.

\begin{figure*}
    \begin{center}
    \includegraphics[height=20cm]{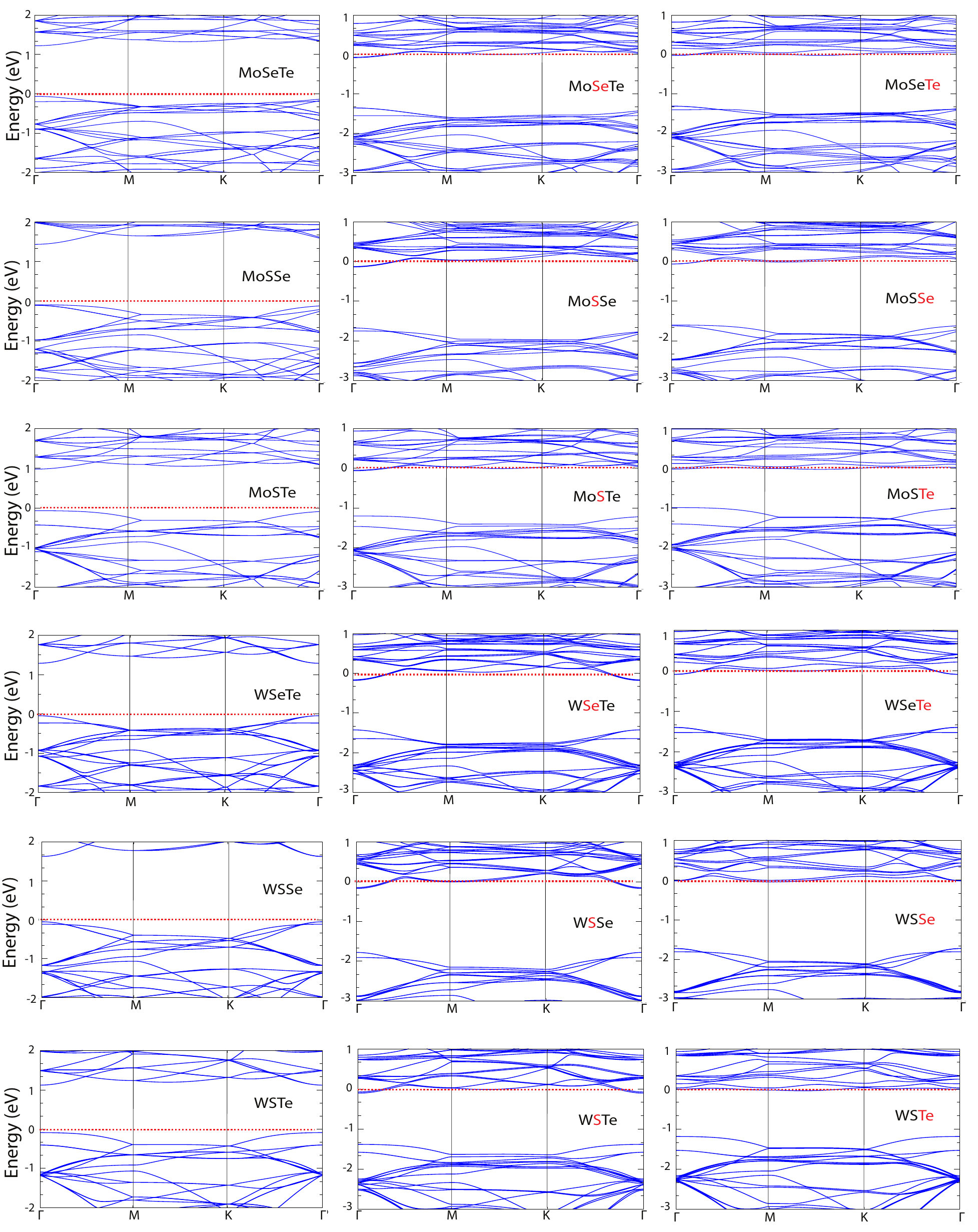}
    \caption{\label{fig:bands}Band structures for Janus monolayers MXY (M=Mo,W; X/Y=S,Se,Te) with $3 \times 3 \times 1$ supercells. Bare structures in left column, lithiated structures in center and right columns.  Red coloring materials naming indicates where Li adatom is adsorbed. Dashed red line indicates the Fermi energy of the system.}
\end{center}
\end{figure*}

We further explored this application to anodes by conducting band structure analysis for the bare and lithiated Janus materials.  All of the examined bare structures are semiconducting.\cite{ersan}  Upon adsorption of a single Li atom, the structures become metallic, as seen in the band structure plots of Fig. \ref{fig:bands}.  Upon Li adsorption at dilute concentrations, the electron transferred from Li atom occupies the conduction band of the underlying Janus TMD structure without altering the band dispersions. This shifts the Fermi level of the system to the conduction band of the Janus TMD layer.  Thus, the structures become more metallic.  This is important because metallic materials are necessary to provide high electrical conductivity in anodes. 

This is also another indication that what we assumed in our CLR ML model was true. At dilute concentrations, the charges transferred from the Li occupy the conduction band and do not alter the electronic structure of the underlying Janus TMD layer. We can safely use the energies of the conduction band edge to relate it to the binding energy of the Li adatom. 

\section{Conclusion}

In this study, we at first examined adsorption and diffusion of a single Li atom on both sides of six Janus TMD systems MXY (M = Mo, W; X/Y = S, Se, Te). We studied charge transfer from the Li atom to the lattice as well as Li-lattice distances.  We examined various Li coverage cases, varying from dilute to full coverage. We included discussions about how these properties affect the binding energy of a Li atom on different regular and Janus TMD surfaces at varying concentrations.  

Based on this analysis, we developed a linear regression model for our concentration-dependent adsorption energy data.  We found that the statistical errors decrease when the data is first sorted into clusters based on supercells, and then regression is performed on each cluster separately.  Overall, this ML model gave us physical insight into the adsorption process. For instance, we found that the energy of the lowest occupied state is a prominent feature in determining adsorption energy of Li on our MX$_{2}$ and Janus MXY materials. Intrinsic dipole of Janus TMDs and concentration-dependent Li-Li interaction terms play a dominant role in predicting the adsorption chemistry at dense doping concentrations. We hope that our results can be used to predict adsorption energies of other adatoms on the same or similar materials.

At last, we focus on analysing these structures as a possible anode in battery applications. An ideal anode material for Li-ion batteries must have a low activation barrier and a high diffusion coefficient.  In these respects, many of the examined Janus materials are comparable or superior to graphene and to regular TMDs.  We found that Li prefers to travel between top-metal sites while passing over metastable hollow sites.  By conducting various stability and performance analysis, we report that the X-sides of Janus structures are suitable for Li adsorption, and that only MoSSe and MoSeTe can be suitable for full coverage of Li atoms on the surface. Multilayer-Li adsorption was hindered due to negative open circuit voltage. Multilayer Janus structures are better suited for battery applications than one-layer Janus structures, as the former experience less volumetric expansion/contraction during discharging/charging process and have higher storage capacity. Finally, efficient electron transport is also a critical property of an electrode.  Janus monolayers transition from semiconducting to metallic upon adsorption of a single Li-ion, which would improve anode conductivity.  Overall, our results imply that the examined Janus structures should perform well as electrodes in Li-ion batteries.   

\section{Associated Content}
The descriptor values ($E_{LUS}$, $E_{Li-TMD(bare)}$, and $E_{Li-Li}$) for the 72 training examples of the ML model are reported in 3 tables in Supporting Information (1). 

A MATLAB code with instructions of calculating the $E_{Li-TMD(bare)}$ descriptor for all TMD structures and python code for applying the machine learning CLR model are provided in Supporting Information (2). 

\section{Author Contributions}
GC conducted the adsorption and diffusion simulations and wrote the manuscript. AI developed the ML model and wrote the manuscript. FE read and edited the manuscript. DC helped analyzing the voltage profiles data. CA designed the study, analyzed the data, conducted the voltage profile simulations and wrote the manuscript.

\section{Acknowledgments}
Part of the calculations have been carried out at UMBC High Performance Computing Facility (HPCF).  This work was supported by the National Science Foundation through Division of Materials Research under NSF DMR-1726213 Grant.

\bibliography{janus_references}

\end{document}


\maketitle
\thispagestyle{fancy}

\begin{abstract}
    In this supporting Information, we present the data for the descriptors used in the proposed machine learning model. Since our dataset included only Li as adatom, we have only one value for the ionization energy ($E_{ion} = 5.39$ eV for Li) \cite{bushaw2007ionization}. Tables S1, S2, and S3 provide the values of the three descriptors defined in the paper, $E_{LUS}$, $E_{Li-TMD(bare)}$, and $E_{Li-Li}$, respectively.   
    
\end{abstract}

\maketitle

\begin{table}[t]
\centering 
    \caption{Data of the energy of lowest unoccupied state ($E_{LUS}$) descriptor (in eV) for all regular/Janus TMD monolayers for coverage ratios for primitive cell [$\theta$(1/3)], $2 \times 2 \times 1$ [$\theta$(1/12)], $3 \times 3 \times 1$ [$\theta$(1/27)], and $4 \times 4 \times 1$ [$\theta$(1/48)] supercells of the cluster-wise linear regression (CLR) model are tabulated. The adsorption site is on top of the metal for all cases. The side of lithiation is also reported.}
    
\footnotesize
  \begin{tabularx}{0.8\textwidth} { 
        | >{\centering\arraybackslash}X 
       | >{\centering\arraybackslash}X 
       | >{\centering\arraybackslash}X 
       | >{\centering\arraybackslash}X 
       | >{\centering\arraybackslash}X
       | >{\centering\arraybackslash}X
       | >{\centering\arraybackslash}X
       | >{\centering\arraybackslash}X |}
        \hline        
        \textbf{TMD} & \textbf{Side} & \boldmath$\theta$\textbf{(1/3)} & \boldmath$\theta$\textbf{(1/12)} & \boldmath$\theta$\textbf{(1/27)}& \boldmath$\theta$\textbf{(1/48)}\\
        \hline

        MoS$_2$ & - &  -4.31731 & -4.31731 & -4.31731 & -4.31731  \\  \hline
         MoSe$_2$ & - &  -3.90168 & -3.90168 & -3.90168 & -3.90168  \\ \hline
         MoTe$_2$ & - &  -3.72932 & -3.72932 & -3.72932 & -3.72932  \\ \hline
         WS$_2$ & - &  -3.94749 & -3.94749 & -3.94749 & -3.94749  \\ \hline
         WSe$_2$ & - &  -3.69454 & -3.69454 & -3.69454 & -3.69454  \\ \hline
         WTe$_2$ & - &  -3.71126 & -3.71126 & -3.71126 & -3.71126  \\ \hline
         MoSSe & S &  -4.46336 & -4.46336 & -4.46336 & -4.46336  \\ \hline
         MoSSe & Se &  -3.78336 & -3.78336 & -3.78336 & -3.78336  \\ \hline
         MoSTe & S &  -4.73547 & -4.73547 & -4.73547 & -4.73547 \\ \hline
         MoSTe & Te &  -3.27547 & -3.27547 & -3.27547 & -3.27547 \\ \hline
         MoSeTe & Se &  -4.24215 & -4.24215 & -4.24215 & -4.24215  \\ \hline
         MoSeTe & Te &  -3.49515 & -3.49515 & -3.49515 & -3.49515  \\ \hline
         WSSe & S &  -4.14052 & -4.14052 & -4.14052 & -4.14052  \\ \hline
         WSSe & Se &  -3.40052 & -3.40052 & -3.40052 & -3.40052  \\ \hline
         WSTe & S &  -4.69068 & -4.69068 & -4.69068 & -4.69068  \\ \hline
         WSTe & Te &  -3.27068 & -3.27068 & -3.27068 & -3.27068  \\ \hline
         WSeTe & Se &  -3.87271 & -3.87271 & -3.87271 & -3.87271  \\ \hline
         WSeTe & Te &  -3.18271 & -3.18271 & -3.18271 & -3.18271  \\ \hline
         
\end{tabularx}
\end{table}

\begin{table}
\centering 
    \caption{Data of the Coulomb interaction energy of a single Li atom with a bare TMD structure ($E_{Li-TMD(bare)}$) descriptor (in eV) for all regular/Janus TMD monolayers for coverage ratios for primitive cell [$\theta$(1/3)], $2 \times 2 \times 1$ [$\theta$(1/12)], $3 \times 3 \times 1$ [$\theta$(1/27)], and $4 \times 4 \times 1$ [$\theta$(1/48)] supercells of the cluster-wise linear regression (CLR) model are tabulated. The adsorption site is on top of the metal for all cases. The side of lithiation is also reported.}
    
\footnotesize
  \begin{tabularx}{0.8\textwidth} { 
       | >{\centering\arraybackslash}X 
       | >{\centering\arraybackslash}X 
       | >{\centering\arraybackslash}X 
       | >{\centering\arraybackslash}X 
       | >{\centering\arraybackslash}X
       | >{\centering\arraybackslash}X
       | >{\centering\arraybackslash}X
       | >{\centering\arraybackslash}X |}
        \hline        
        \textbf{TMD} & \textbf{Side} & \boldmath$\theta$\textbf{(1/3)} & \boldmath$\theta$\textbf{(1/12)} & \boldmath$\theta$\textbf{(1/27)}& \boldmath$\theta$\textbf{(1/48)}\\
        \hline
       
         MoS$_2$ & - &  -0.00110 & -0.00110 & -0.00110 & -0.00110  \\  \hline
         MoSe$_2$ & - & -0.00120 & -0.00120 & -0.00120 & -0.00120  \\ \hline
         MoTe$_2$ & - &  -0.00065 & -0.00065 & -0.00065 & -0.00065  \\ \hline
         WS$_2$ & - &  -0.00310 & -0.00310 & -0.00310 & -0.00310  \\ \hline
         WSe$_2$ & - &  -0.00024 & -0.00024 & -0.00024 & -0.00024  \\ \hline
         WTe$_2$ & - & -0.00036 & -0.00036 & -0.00036 & -0.00036  \\ \hline
         MoSSe & S & -1.33570 & -1.33570 & -1.33570 & -1.33570  \\ \hline
         MoSSe & Se &  1.33330 & 1.33330 & 1.33330 & 1.33330  \\ \hline
         MoSTe & S &  -3.91630 & -3.91630 & -3.91630 & -3.91630  \\ \hline
         MoSTe & Te &  3.91410 & 3.91410 & 3.91410 & 3.91410 \\ \hline
         MoSeTe & Se &  -2.40680 & -2.40680 & -2.40680 & -2.40680  \\ \hline
         MoSeTe & Te & 2.40510 & 2.40510 & 2.40510 & 2.40510  \\ \hline
         WSSe & S &  -1.83160 & -1.83160 & -1.83160 & -1.83160  \\ \hline
         WSSe & Se &  1.82890 & 1.82890 & 1.82890 & 1.82890  \\ \hline
         WSTe & S &  -4.32230 & -4.32230 & -4.32230 & -4.32230  \\ \hline
         WSTe & Te &  4.31970 & 4.31970 & 4.31970 & 4.31970  \\ \hline
         WSeTe & Se & -2.59010 & -2.59010 & -2.59010 & -2.59010  \\ \hline
         WSeTe & Te &  2.58820 & 2.58820 & 2.58820 & 2.58820  \\ \hline
         
\end{tabularx}
\end{table}
\clearpage
\begin{table}
\centering 
    \caption{Data of the Li-Li interaction energy ($E_{Li-Li}$) descriptor (in eV) for all regular/Janus TMD monolayers for coverage ratios for primitive cell [$\theta$(1/3)], $2 \times 2 \times 1$ [$\theta$(1/12)], $3 \times 3 \times 1$ [$\theta$(1/27)], and $4 \times 4 \times 1$ [$\theta$(1/48)] supercells of the cluster-wise linear regression (CLR) model are tabulated. The adsorption site is on top of the metal for all cases. The side of lithiation is also reported.}
    
\footnotesize
  \begin{tabularx}{0.8\textwidth} { 
       | >{\centering\arraybackslash}X 
       | >{\centering\arraybackslash}X 
       | >{\centering\arraybackslash}X 
       | >{\centering\arraybackslash}X 
       | >{\centering\arraybackslash}X
       | >{\centering\arraybackslash}X
       | >{\centering\arraybackslash}X
       | >{\centering\arraybackslash}X |}
        \hline        
        \textbf{TMD} & \textbf{Side} & \boldmath$\theta$\textbf{(1/3)} & \boldmath$\theta$\textbf{(1/12)} & \boldmath$\theta$\textbf{(1/27)}& \boldmath$\theta$\textbf{(1/48)}\\
        \hline
       
         MoS$_2$ & - &  5.82719 & 4.38995 & 3.90081 & 3.94406  \\  \hline
         MoSe$_2$ & - &  5.80762 & 4.42304 & 3.95263 & 3.99413  \\ \hline
         MoTe$_2$ & - & 5.77492 & 4.47550 & 4.03497 & 4.07376  \\ \hline
         WS$_2$ & - & 5.82872 & 4.38729 & 3.89664 & 3.94005  \\ 
         \hline
         WSe$_2$ & - & 5.80614 & 4.42549 & 3.95645 & 3.99782  \\ \hline
         WTe$_2$ & - & 5.77220 & 4.47974 & 4.04152 & 4.08020  \\ \hline
         MoSSe & S & 5.81808 & 4.40556 & 3.92521 & 3.96766  \\ 
         \hline
         MoSSe & Se & 5.81808 & 4.40556 & 3.92521 & 3.96766  \\ \hline
         MoSTe & S & 5.80320 & 4.43032 & 3.97641 & 4.00513  \\ 
         \hline
         MoSTe & Te & 5.80320 & 4.43032 & 3.97641 & 4.00513 \\ 
         \hline
         MoSeTe & Se & 5.79306 & 4.44675 & 3.99027 & 4.03006  \\ \hline
         MoSeTe & Te & 5.79306 & 4.44675 & 3.99027 & 4.03006  \\ \hline
         WSSe & S & 5.81960 & 4.40300 & 3.92120 & 3.96379  \\ 
         \hline
         WSSe & Se & 5.81960 & 4.40300 & 3.92120 & 3.96379  \\ 
         \hline
         WSTe & S & 5.80320 & 4.43032 & 3.97641 & 4.00513  \\ 
         \hline
         WSTe & Te & 5.80320 & 4.43032 & 3.97641 & 4.00513  \\ 
         \hline
         WSeTe & Se & 5.79306 & 4.44675 & 3.99027 & 4.03006  \\ 
         \hline
         WSeTe & Te & 5.79306 & 4.44675 & 3.99027 & 4.03006  \\ 
         \hline
         
  \end{tabularx}
\end{table}

\bibliography{janus_references.bib}